\documentclass[usletter,12pt]{article}
\pdfoutput=1 % if your are submitting a pdflatex (i.e. if you have
\usepackage{jheppub} % for details on the use of the package, please
\usepackage[T1]{fontenc} % if needed

\usepackage{amsmath,amsfonts,epsf}
\usepackage{amssymb}
\usepackage{graphicx}
\usepackage{grffile}
\usepackage{dsfont}
\usepackage{hyperref}
\input epsf

\def\bC {\mathds{C}}

\newcommand{\A}{{\alpha}}
\newcommand{\B}{{\beta}}
\newcommand{\C}{{\gamma}}
\newcommand{\D}{{\delta}}

\newcommand{\pd}{\partial}

\newcommand{\vev}[1]{{\left< {#1} \right>}}

\numberwithin{equation}{section}

\begin{document}

\title{Constraining conformal field theories with a higher spin symmetry in $d > 3$ dimensions}
\author[]{Vasyl Alba,}
\author[]{Kenan Diab}

\affiliation[]{Department of Physics 
\\ 
Jadwin Hall, Princeton University,\\
Princeton, NJ 08544 USA}

\emailAdd{valba@princeton.edu}
\emailAdd{kdiab@princeton.edu}

\abstract{We study unitary conformal field theories with a unique stress tensor and at least one
higher-spin conserved current in $d>3$ dimensions. We prove that every such theory contains an
infinite number of higher-spin conserved currents of arbitrarily high spin, and that Ward identities
generated by the conserved charges of these currents imply that the correlators of the stress tensor
and the conserved currents of the theory must coincide with one of the following three
possibilities: a) a theory of $n$ free bosons (for some integer $n$), b) a theory of $n$ free
fermions, or c) a theory of $n$ $\frac{d-2}{2}$-forms. For $d$ even, all three structures exist, 
but for $d$ odd, it may be the case that the third structure (c) does not; if it does exist, it is
unclear what theory, if any, realizes it. This is a generalization of the result proved in three
dimensions by Maldacena and Zhiboedov \cite{Maldacena:2011jn}. This paper supersedes the previous
paper by the authors \cite{Alba:2013yda}.}

\arxivnumber{1510.xxxx}
\maketitle

\section*{Introduction}\label{introduction}
\addcontentsline{toc}{section}{Introduction}
Characterizing the theories dual to Vasiliev's higher-spin gauge theories in anti
de-Sitter space\cite{Konstein:2000bi}\cite{Vasiliev:2003ev}\cite{Vasiliev:2004qz} under the AdS/CFT
correspondence\cite{Maldacena:1997re}\cite{Gubser:1998bc}\cite{Witten:1998qj} has been a topic of
active research for over ten years, starting from the conjecture of Klebanov and Polyakov that
Vasiliev's theory in four dimensions is dual to the critical $O(N)$ vector model in three 
dimensions\cite{Klebanov:2002ja}\cite{Sezgin:2003pt}. Under general principles of AdS/CFT, we
expect that the conformal field theory duals to Vasiliev's theories (when given appropriate boundary
conditions) should also have higher-spin symmetry, so it is natural to try to classify all
higher-spin conformal field theories. In the case of CFT's in three dimensions, this task has
already been accomplished by Maldacena and Zhiboedov\cite{Maldacena:2011jn}, who showed that unitary
conformal field theories with a unique stress tensor and a higher-spin current are essentially free
in three dimensions. This can be viewed as an analogue of the Coleman-Mandula
theorem\cite{Coleman:1967ad}\cite{Haag:1974qh}, which states that the maximum spacetime symmetry of
theories with a nontrivial S-matrix is the super-Poincare group, along with any internal symmetries
whose charges are Lorentz-invariant quantum numbers (i.e.~are scalars with respect to the spacetime
symmetry group). 

In this paper, we will prove an analogue of the Coleman-Mandula theorem for generic conformal field
theories in all dimensions greater than three. We will show that in any conformal field
theory that (a) satisfies the unitary bound for operator dimensions, (b) satisfies the cluster
decomposition axiom, (c) contains a symmetric conserved current of spin larger than $2$, and 
(d) has a unique stress tensor in $d>3$ dimensions, all correlation functions of symmetric currents
of the theory are equal to the correlation functions of one of the following three theories - either
the theory of $n$ free bosons (for some integer $n$), a theory of $n$ free fermions, or a theory of
$n$ free $\frac{d-2}{2}$-forms. 

Note that in odd dimensions, the free $\frac{d-2}{2}$-form does not exist, and the status of our
result is somewhat complicated. We do not show that there exists any solution to the conformal Ward
identities that corresponds to this possibility in odd dimensions, although we do show that if one exists,
it is unique. For every odd dimension $d \ge 7$, we know that an infinite tower of higher-spin currents
must be present \cite{Boulanger:2013zza}, but in $d=5$, it may be the case that there are not infinitely
many higher spin currents. Assuming that the solution exists and there are an infinite number of
higher spin currents, we show that the correlation functions of the conserved currents of the theory
may be understood as the analytic continuation of the correlation functions of the currents of the
even-dimensional free $\frac{d-2}{2}$-form theory to odd dimensions. Then, even under all these
assumptions, we do not show that there exists any conformal field theory that realizes this
solution.  That is, it is possible that this structure may have no good microscopic interpretation
for other reasons. For example, in odd dimensions it could be possible that some correlation
function of some operators is not consistent with the operator product expansion in the sense that it
cannot be decomposed in a sum over conformal blocks with non-negative coefficients (i.e. consistent with 
unitarity\footnote{There is an example of this phenomenon. If one considers a theory of $N$ scalar
fields $\phi_i$ and computes the four-point function of the operator $\phi^2 = \sum_i \phi_i\phi_i$,
it turns out that $N$ should be greater then 1, otherwise the theory is nonunitary.}). Such questions
are not explored in this work.

Furthermore, we note that a recent paper by Boulanger, Ponomarev, Skvortsov, and Taronna
\cite{Boulanger:2013zza} strongly indicates that all the algebras of higher-spin charges that are
consistent with conformal symmetry are not only Lie algebras but associative. Hence, they are all
reproduced by the universal enveloping construction of \cite{Boulanger:2011se} with the conclusion
that any such algebra must contain a symmetric higher-spin current. This implies that our result
should be true even after relaxing our assumption that the higher-spin current is symmetric. The
argument is structured as follows:

\begin{description}
\item[In section \ref{lightcone-limits},]we will present the main technical tool of the paper: we
will define a particular limit of three-point functions of symmetric conserved currents called
\textit{lightcone limits}. We will show that such correlation functions behave essentially like
correlation functions of a free theory in these limits, enabling us to translate complicated Ward
identities of the full theory into simpler ones involving only free field correlators. We will also
compute the Fourier transformation of these correlation functions; this will ultimately allow us to
simplify certain Ward identities into easily-analyzed polynomial equations. 
\end{description}

The rest of the paper will then carry out proof of our main statement. The steps are as follows:
\begin{description}
\item[In section \ref{charge-conservation-identities},]we will solve the Ward identity arising from
the action of the charge $Q_s$ arising from a spin $s$ current $j_s$ on the correlator 
$\vev{j_2j_2j_s}$ in the lightcone limit, where $j_2$ is the stress tensor. We will show
that the only possible solution is given by the free-field solution. This implies the existence of
infinitely many conserved currents of arbitrarily high spin,\footnote{The fact that the existence of
a higher-spin current implies the existence of infinitely many other higher-spin currents has been
proven before in the four-dimensional case \cite{Komargodski:2012ek} under the additional
assumptions that the theory flows to a theory with a well defined S-matrix in the infrared, that the
correlation function $\vev{j_2j_2j_s} \neq 0$, and that the scattering amplitudes of the theory
have a certain scaling behavior. This statement was also proven for $d \neq 4,5$ in 
\cite{Boulanger:2013zza} by classifying all the higher-spin algebras in all dimensions other than 
$4$ and $5$. We give a proof for the
sake of completeness, and also because our techniques differ from those two papers.} thereby giving
rise to infinitely many charge conservation laws which powerfully constrain the theory. 
\item[In section \ref{bilocal-fields},]we will construct certain quasi-bilocal fields which roughly
behave like products of free fields in the lightcone limit, yet are defined for any CFT. We will
establish that all the higher-spin charges (whose existence was proven in the previous step) act on
these quasi-bilocals in a particularly simple way.
\item[In section \ref{qb-correlators},]we will translate the action of the higher-spin charges on
the quasi-bilocals into constraints on correlation functions of the quasi-bilocals. We will then
show that these constraints are so powerful that they totally fix every correlation function of the
quasi-bilocals to agree with the corresponding correlation function of a particular biprimary operator
in free field theory on the lightcone.
\item[In section \ref{constraining-correlators},]we show how the quasi-bilocal correlation functions can
be used to prove that the three-point function of the stress tensor must be equal to the three-point
function of either the free boson, the free fermion, or the free $\frac{d-2}{2}$-form, even away from the
lightcone limit. This is then used to recursively constrain every correlation function of the CFT
to be equal to the corresponding correlation function in the free theory, finishing the proof.
\end{description}
This strategy is similar to the argument in the three-dimensional case given in \cite{Maldacena:2011jn}. 
There are two main differences between the three-dimensional case and the higher-dimensional cases
that we must account for:

First, the Lorentz group in $d>3$ admits asymmetric representations, but the
three-dimensional Lorentz group does not. By asymmetric, we mean that a current
$J_{\mu_1\dots\mu_n}$ is not invariant with respect to interchange of its indices. For example, in
in the standard $(j_1,j_2)$ classification of representations of the four-dimensional Lorentz group
induced from the isomorphism of Lie algebras $\mathfrak{so}(3,1)_\bC \cong \mathfrak{sl}(2,\bC)
\oplus \mathfrak{sl}(2,\bC)$, these are the representations with $j_1 \neq j_2$. The existence of
these representations means that many more structures are possible in $d>3$ dimensions than in three
dimensions (the asymmetric structures), and so many more coefficients have to be constrained in
order to solve the Ward identities. We restrict our attention to Ward identities arising from the
action of a symmetric charge to correlation functions of only symmetric currents; we will then show
that asymmetric structures cannot appear in these Ward identities, making the exact solution of the
identities possible.

Second, the space of possible correlation functions consistent with conformal symmetry is larger in
$d>3$ dimensions than in three dimensions. For example, consider the three-point function of the
stress tensor $\vev{j_2j_2j_2}$. It has long been known (see,
e.g.~\cite{Stanev:1988ft}\cite{Osborn:1993cr}\cite{Stanev:2012nq}\cite{Zhiboedov:2012bm}) that this
correlation function factorizes into three structures in $d>3$ dimensions, as opposed to only two
structures in three dimensions (ignoring a parity-violating structure which is eliminated in three
dimensions by the higher-spin symmetry). These three structures correspond to the correlation
functions that appear in the theories of free bosons, free fermions, and free $\frac{d-2}{2}$-forms. We
will show that even though more structures are possible in four dimensions and higher, the Ward
identities we need can still be solved.

We note that our work is related to a paper by Stanev \cite{Stanev:2013qra}, in
which the four, five, and six-point correlation functions of the stress tensor were
constrained in CFT's with a higher spin current in four dimensions. It was also shown that the pole
structure of the general $n$-point function of the stress tensor coincides with that of a free field
theory. Though this paper reaches the same conclusions, we do not make the
rationality assumption \cite{Nikolov:2000pm} of that paper.

Finally, while this paper was being prepared, the paper \cite{Friedan:2015xea} appeared in which
they showed that unitary ``Cauchy conformal fields'', which are fields that satisfy a certain first-order
differential equation, are free in the sense that their correlation functions factorize on the
2-point function. Their result may be understood as establishing a similar result that applies even to
certain fields which are not symmetric traceless, which we say nothing about.
\section{Definition of the lightcone limits}\label{lightcone-limits} 
The fundamental technical tool we need to extend into four dimensions and higher is the
\textit{lightcone limit}. In order to constrain the correlation functions of the theory to be equal
to free field correlators, we will show that the three-point function of the $\vev{j_2j_2j_2}$ must
be equal to $\vev{j_2j_2j_2}$ for a free boson, a free fermion, or a free $\frac{d-2}{2}$-form field
- it cannot be some linear combination of these three structures. To this end, it will be helpful to
split up the Ward identities of the theory into three different identities, each of which involves
only one of the three structures separately. To do this, we will need to somehow project all the
three-point functions of the theory into these three sectors. The lightcone limits accomplish this
task.

Before defining the lightcone limits, we will set up some notation. As in \cite{Maldacena:2011jn}, 
we are writing the flat space metric $ds^2 = dx^+dx^- + d\vec{y}^2$ and contracting each current
with lightline polarization vectors whose only nonzero component is in the minus direction: $j_s
\equiv J_{\mu_1\dots\mu_s}\epsilon^{\mu_1}\dots\epsilon^{\mu_s} = J_{--\dots-}$. We will also
denote $\pd_1 \equiv \pd/\pd x_1^-$ and similarly for $\pd_2$ and $\pd_3$. Thus, in all
expressions where indices are suppressed, those indices are taken to be minus indices.
There are two things we will establish:
\begin{enumerate}
\item We need to define an appropriate limit for each of the three cases, which, when applied to a
three-point function of conserved currents $\vev{\underline{j_{s_1}j_{s_2}}j_{s_3}}$, yields an
expression proportional to an appropriate correlator of the free field theory. For example, in the
bosonic case where all the currents are symmetric, we would like the lightcone limit to give us
$\pd_1^{s_1}\pd_2^{s_2}\vev{\phi\phi^*j_{s_3}}_{\text{free}}$.
\item Second, we need to explicitly compute the free field correlator which we obtain from the
lightcone limits. In the bosonic case where all currents are symmetric, this would mean that we
need to compute the three-point function $\vev{\phi\phi^*j_{s_3}}$ in the free theory.
\end{enumerate}

For the first task, we claim that the desired lightcone limits are:
\begin{align}
\vev{\underline{j_{s_1} j_{s_2}}_b j_{s_3}} &\equiv \lim_{|y_{12}|\rightarrow 0} |y_{12}|^{d-2} 
\lim_{x^+_{12}\rightarrow 0} \vev{j_{s_1} j_{s_2} j_{s_3}} \propto
\pd_1^{s_1}\pd_2^{s_2}\vev{\phi\phi^*j_{s_3}}_{\text{free}} \label{e:bosonic-lcl}\\
\vev{\underline{j_{s_1} j_{s_2}}_f j_{s_3}} &\equiv \lim_{|y_{12}|\rightarrow 0} |y_{12}|^d
\lim_{x^+_{12}\rightarrow 0} \frac{1}{x^+_{12}} \vev{j_{s_1} j_{s_2} j_{s_3}} \propto
\pd_1^{s_1-1}\pd_2^{s_2-1}\vev{\psi\C_-\bar{\psi}j_{s_3}}_{\text{free}} \label{e:fermionic-lcl}\\
\vev{\underline{j_{s_1} j_{s_2}}_t j_{s_3}} &\equiv \lim_{|y_{12}|\rightarrow 0} |y_{12}|^{d+2} 
\lim_{x^+_{12}\rightarrow 0} \frac{1}{(x^{+}_{12})^2}\vev{j_{s_1} j_{s_2} j_{s_3}} \propto
\pd_1^{s_1-2}\pd_2^{s_2-2}\vev{F_{-\{\alpha\}}F_{-\{\alpha\}}j_{s_3}}_{\text{free}} \label{e:vector-lcl}
\end{align}
Here, the subscript \textit{b, f,} and \textit{t} denote the bosonic, fermionic, and tensor
lightcone limits. $\phi$ is a free boson, $\psi$ is a free fermion, and $F$ is the 
field tensor for a free $\frac{d-2}{2}$-form field; the repeated $\{\alpha\}$ indices indicate
Einstein summation over all other indices. For example, in four dimensions, the ``tensor''
structure is just the ordinary free Maxwell field. For conciseness, we will often refer to the free
$\frac{d-2}{2}$-form field as simply the ``tensor field'' or the ``tensor structure''. Again, we
emphasize that in odd dimensions, the free $\frac{d-2}{2}$-form field does not exist. In odd
dimensions, our claim is that the only possible structure with the scaling behavior captured by the
tensor lightcone limit is the one which coincides with the naive analytic continuation of the
correlation functions of the free $\frac{d-2}{2}$ form field to odd $d$.

The justification for the first two equations comes from the generating functions obtained in
\cite{Stanev:2012nq}\cite{Zhiboedov:2012bm}; in those references, the three-point functions for
correlation functions of conserved currents with $y_{12}$ and $x_{12}^+$ dependence of those types
was uniquely characterized, and so taking the limit of those expressions as indicated gives us the
claimed result. In the tensor case, \cite{Zhiboedov:2012bm} did not find a unique structure, but
rather, a one-parameter family of possible structures. Nevertheless, all possible structures
actually coincide in the lightcone limit, as is proven in appendix \ref{vector-uniqueness}. 

We note that parity-violating structures cannot appear after taking these lightcone limits.
This is because the all-minus component of every parity violating structure allowed by 
conformal invariance in $d>3$ dimensions is identically zero. 
To see this, observe that all parity-violating structures for three-point functions
consistent with conformal symmetry must have exactly one $\epsilon_{\mu_1\mu_2\dots\mu_d}$ tensor
contracted with polarization vectors and differences in coordinates. Only two of these differences
are independent of each other, and all polarization vectors in the all-minus components are set to
be equal. Thus, there are only three unique objects that can be contracted with the $\epsilon$
tensor, but we need $d$ unique objects to obtain a nonzero contraction. Thus, all parity-violating
structures have all-minus components equal to zero.

Later in our argument, we will need expressions for the Fourier transformation of the
lightcone-limit three point function of two free fields and a spin $s$ current with respect to the
variables $x_1^-$ and $x_2^-$ in the theories of a free boson, a free fermion, and a free
$\frac{d-2}{2}$-form field. The computation for each of the three cases is straightforward and is
given explicitly in appendix \ref{fourier-transforms}. The results are as follows:
\begin{align}
F^b_s &\equiv \vev{\underline{\phi\phi^*}j_s} \propto
(p^+_2)^s{_2}F_1\left(2-\frac{d}{2}-s,-s,\frac{d}{2}-1,p^+_1/p^+_2\right)\\
F^f_s &\equiv \vev{\underline{\psi\gamma_-\bar{\psi}}j_s} \propto
(p^+_2)^{s-1}{_2}F_1\left(1-\frac{d}{2}-s,-s,\frac{d}{2},p^+_1/p^+_2\right)\\
F^b_s &\equiv \vev{\underline{F_{-\{\alpha\}}F_{-\{\alpha\}}}j_s} \propto
(p^+_2)^{s-2}{_2}F_1\left(-\frac{d}{2}-s,-s,\frac{d}{2}+1,p^+_1/p^+_2\right)
\label{e:hypergeometric}
\end{align}
Here, ${_2}F_1$ is the hypergeometric function, and the proportionality sign in each formula
indicates that we have omitted an overall nonsingular function which we are not interested in. That
they are indeed nonsingular is also proven in appendix \ref{fourier-transforms}.

Before continuing, we emphasize that the three lightcone limits we have defined do not
cover all possible lightcone behaviors which can be realized in a conformal field theory. We define
only these three limits because one crucial step in our proof is to constrain the three-point
function of the stress tensor $\vev{j_2j_2j_2}$, which has only these three scaling behaviors. 

Furthermore, though we have discussed only symmetric currents, one could hope that similar
expressions could be generated for asymmetric currents - that is, lightcone limits of correlation
functions of asymmetric currents are generated by one of the three free field theories discussed
here. Unfortunately, running the same argument in \cite{Zhiboedov:2012bm} fails in the case of
asymmetric currents in multiple ways. Consider the current $\vev{j_2j_s\bar{j}_s}$, where $j_s$ is
some asymmetric current and $\bar{j}_s$ is its conjugate. To determine how such a correlator could
behave the lightcone limit, one could write out all the allowed conformally invariant structures
consistent with the spin of the fields, and seeing how each one behaves in the lightcone limits.
Unlike the symmetric cases, one finds that in the lightcone limit many independent structures exist, 
and these structures behave differently depending on which pair of coordinates we take the
lightcone limit. To put it another way, for a symmetric current $s$, one has the decomposition:
\begin{equation}
\vev{j_2j_sj_s} = \sum_{j \in \{b,f,t\}} \vev{j_2j_sj_s}_j
\end{equation}
where the superscript $j$ denotes the result after taking corresponding lightcone limit in any of
the three pairs of coordinates (all of which yield the same result), and the corresponding
structures can be understood as arising from some free theory. In the case of asymmetric $j_s$,
this instead becomes a triple sum
\begin{equation}
\vev{j_2j_s\bar{j}_s} = \sum_{j,k,l \in \{b,f,t\}} \vev{j_2j_s\bar{j}_s}_{(j,k,l)}
\end{equation}
where each sum corresponds to taking a lightcone limit in each of the three different pairs of
coordinates, and we do not know how to interpret the independent structures in terms of a free field
theory. This tells us that for asymmetric currents, the lightcone limit no longer achieves its
original goal of helping us split up the Ward identities into three identities which can be analyzed
independently; each independent structure could affect multiple different Ward identities.
Again, we emphasize that this does not exclude the possibility of a different lightcone limit reducing
the correlators of asymmetric currents to those of some other free theory. It simply means that our
techniques are not sufficient to constrain correlation functions involving asymmetric currents, so
we will restrict our attention to correlation functions that involve only symmetric currents.

\section{Charge conservation identities}\label{charge-conservation-identities}
We will now use the results of the previous section to prove that every CFT with a higher-spin
current contains infinitely many higher-spin currents of arbitrarily high (even) spin. 
We note that this result was proven in a different way in \cite{Boulanger:2013zza} for all
dimensions other than $d=4$ and $d=5$, wherein they showed that there is a unique higher-spin 
algebra in $d \neq 4,5$ and showed that they all infinitely many higher-spin currents.
The discussion below is a different proof of this statement based on analysis of the 
constraints that conservation of the higher-spin charge imposes, and the techniques we develop here
will be used later. As before, we treat the bosonic, fermionic, and tensor cases separately.

Before beginning, we will tabulate a few results about commutation relations that we will use freely
throughout from this section onwards. Their proofs are identical to those in
\cite{Maldacena:2011jn}, and are therefore omitted:
\begin{enumerate}
\item If a current $j'$ appears (possibly with some number of derivatives) in the commutator
$[Q_s,j]$, then $j$ appears in $[Q_{s},j']$. 
\item Three-point functions of a current with odd spin with two identical currents of even spin are
zero: $\vev{j_sj_sj_{s'}} = 0$ if $s$ is even and $s'$ is odd.
\item The commutator of a symmetric current with a charge built from another symmetric current
contains only symmetric currents and their derivatives:
\begin{equation}
[Q_s,j_{s'}] = \sum_{s'' = \max[s'-s+1,0]}^{s'+s-1} \alpha_{s,s',s''}\pd^{s'+s-1-s''}j_{s''}
\label{e:commutator-expansion}
\end{equation}
The proof of this statement requires an additional step since one needs to exclude asymmetric
currents contracted with invariant symbols like the $\epsilon$ tensor. For example, consider what
structures could appear in $[Q_2,j_2]$ in four dimensions. In $SU(2)$ indices, this object has
three dotted and three undotted spinor indices, so one could imagine that a structure like
$\epsilon_{ab}j^{abcde\dot{c}\dot{d}\dot{e}}$ could appear in $[Q_2,j_2]$. However, $[Q_2,j_2]$ has
conformal dimension $5$, and the unitarity bound constrains the current $j$, which transforms in the
$(5/2,3/2)$ representation, to have conformal dimension at least $d-2+s = 6$, which is impossible.
The proof for a general commutator $[Q_s,j_{s'}]$ follows in an identical manner.
\item $[Q_s,j_2]$ contains $\pd j_s$. This was actually proven for all dimensions in appendix A 
of \cite{Maldacena:2011jn}. Item $1$ then implies that $[Q_s,j_s]$ contains $\pd^{2s-3}j_2$.
\end{enumerate}
In these statements, we are implicitly ignoring the possibility of parity violating structures. For
example, the three-point function $\vev{221}$, which is related to the $U(1)$ gravitational anomaly, 
may not be zero in a parity violating theory. As mentioned in section \ref{lightcone-limits},
however, the all-minus components of every parity-violating structure consistent with conformal
symmetry is identically zero, so they will not appear in any of our identities here.

Let's start with the bosonic case.
Consider the charge conservation identity arising from the action of $Q_s$ on $\vev{\underline{22}_bs}$:
\begin{equation}
0 = \vev{\underline{[Q_s,2]2}_bs} + \vev{\underline{2[Q_s,2]}_bs} + \vev{\underline{22}_b[Q_s,s]} 
\end{equation}
If $s$ is symmetric, we may use the general commutation relation (\ref{e:commutator-expansion}) and
the lightcone limit (\ref{e:bosonic-lcl}) to expand this equation out in terms of free field
correlators:
\begin{equation}
0 =
\partial_1^2\partial_2^2\left(\C(\partial_1^{s-1}+(-1)^s\partial_2^{s-1})\vev{\underline{\phi\phi^*}s}_{free}
+ \sum_{2\le k < 2s-1\text{ even}}
\tilde{\alpha_k}\partial_3^{2s-1-k}\vev{\underline{\phi\phi^*}k}_{free}\right)
\label{e:bosonic-charge-identity}
\end{equation}
Note that the sum over $k$ is restricted to even currents since $\vev{22k} = 0$ for odd $k$. In
addition, the fact that the coefficient in front of the $\pd^{s-1}_2$ term is constrained to be
$(-1)^s$ times the coefficient for the $\pd^{s-1}_1$ term arises from the symmetry of
$\vev{\phi(x_1)\phi^*(x_2)j_s(x_3)}$ under interchange of $x_1$ and $x_2$.

Now, we apply our Fourier space expressions for the three-point functions given in section 
\ref{lightcone-limits}. In the Fourier transformed variables, derivatives along the
minus direction turn into multiplication by the momenta in the plus direction. 
After ``cancelling out'' the overall derivatives, which just yields an overall factor of
$(p_1^+)^2(p_2^+)^2$, the relevant equation is:
\begin{equation}
0 = \gamma((p_1^+)^{s-1}+(-1)^s(p_2^+)^{s-1})F_s(p_1^+, p_2^+) +
\sum_{2\le k<2s-1\text{ even}} \tilde{\A}_k (p_1^++p_2^+)^{2s-1-k}
F_k(p_1^+,p_2^+)
\label{e:fourier-bosonic-charge-identity}
\end{equation}
The solution of (\ref{e:fourier-bosonic-charge-identity}) is not easy to obtain by direct calculation. 
We can make two helpful observations, however. First, not all coefficients can
be zero. This is because we know $2$ appears in $[Q_s,s]$, so at least $\tilde{\alpha}_2$ is not
zero. Second, we know that the free boson exists (and is a CFT with higher spin symmetry), and
therefore, the coefficients one obtains from that theory would exactly solve this equation. We will
show that this solution is unique.

Suppose we have two sets of coefficients $(\C,\{\tilde{\A}_k\})$ and $(\C',
\{\tilde{\B}_k\})$ that solve this equation. First, suppose $\C \neq 0$ and $\C' \neq 0$. Then, 
we can normalize the coefficients so that $\C = \C'$ are equal for the two
solutions. Then, subtract the two solutions from each other so that the $\C$ terms vanish. If we
evaluate the result at some arbitrary nonzero value of $p_2^+$, we may absorb all overall $p_2^+$
factors into the coefficients and re-express the equation as a polynomial identity in a
single variable $z \equiv p_1^+/p_2^+$: 
\begin{equation}
0 = \sum_{2\le k<2s-1\text{ even}}
\tilde{\D}_k(1+z)^{2s-1-k}{_2}F_1(2-\frac{d}{2}-k,-k,\frac{d}{2}-1,-z)
\end{equation} 
Then, the entire right hand side is divisible by $1+z$ since $s$ is even, so we may divide both
sides by $1+z$. Setting $z=-1$, since ${_2}F_1(a,a,1,1) \neq 0$ for all negative
half-integers $a$, we conclude that
$\tilde{\delta}_{2s-2} = 0$. Then, the entire right hand side is proportional to $(1+z)^2$, so we
may divide it out. Then, setting $z=-1$ again, we find $\tilde{\delta}_{2s-4} = 0$. Repeating this
procedure, we conclude that all coefficients are zero, and therefore, that the two solutions are
identical. On the other hand, suppose one of the solutions has $\C = 0$. Then, the same argument
establishes that all the coefficients $\tilde{\A}_k$ are zero. As noted earlier, however, the 
trivial solution is disallowed. Therefore, the solution is unique and coincide with one for free
boson. Thus, we have infinitely many even conserved currents, as desired.

In the fermionic case, precisely the same analysis works. The action of $Q_s$ on
$\vev{\underline{22}_fs}$ for symmetric $s$ leads to
\begin{equation}
0 = \partial_1^2\partial_2^2 \Big(\C(\partial_1^{s-2}+(-1)^{s-1}\partial_2^{s-2})\vev{\psi\bar\psi s} +
\sum_{2\le k < 2s-2\text{ even}} \tilde{\alpha}^k\partial_3^{2s-2-k}\vev{\psi\bar\psi k}\Big),
\end{equation}
Then, converting this expression to form factors and running the same analysis from the
bosonic case verbatim establishes that the unique solution to this equation is the one arising in
the theory of a free fermion.

In the tensor case, the argument again passes through exactly as before, except for two subtleties:

First, unlike in the bosonic and fermionic case, we do not have unique expressions for the
three-point functions of currents with the tensor-type coordinate dependence, so this only
demonstrates that the free-field solution is an admissible solution, but not necessarily the unique
solution. Nevertheless, in the lightcone limit, all possible structures for three-point functions
coincide with the free-field answer.\footnote{Actually, we proved that correlators of the form
$\vev{22s}$ have a unique tensor structure even away from the lightcone limit. The proof, however,
is very technical, and it is given in appendix \ref{uniqueness-of-22s}.} This was proven in appendix
\ref{vector-uniqueness}. 

Second, there may not exist a solution to the Ward identities in odd dimensions, because the free
$\frac{d-2}{2}$-form does not exist in odd dimensions. However, if any solution exists, our
argument shows that it is unique. In $d \ge 7$, it is known that there is a unique higher-spin
algebra containing the tower of higher-spin currents described in the bosonic and fermionic cases
\cite{Boulanger:2013zza}. In $d=5$, our technique shows that if there is a solution
for the Ward identity in the tensor lightcone limit, then it is unique.  We do not prove, however,
that there is an infinite tower of higher spin currents or that there is exactly one current of
every spin. Finite dimensional representations would be inconsistent with unitarity. We do not
explore this question further in this work.  Henceforth, we assume that our theory does indeed
contain the infinite tower of higher-spin currents necessary for our analysis.

\section{Quasi-bilocal fields: basic properties}\label{bilocal-fields}
In this section, we will define a set of quasi-bilocal operators, one for each of the three
lightcone limits, and characterize the charge conservation identities arising from the action of the
higher-spin currents. As we will explain in section \ref{qb-correlators}, these charge conservation
identities will turn out to be so constraining that the correlation functions of the quasi-bilocal
operators are totally fixed. This will then enable us to recursively generate all the correlation
functions of the theory and prove that the three-point function of the stress tensor can exhibit
only one of the three possible structures allowed by conformal symmetry. As in the
three-dimensional case, we define the quasi-bilocal operators on the lightcone as operator product
expansions of the stress tensor with derivatives ``integrated out'':
\begin{align}
\underline{22}_b&=\partial_1^2\partial_2^2B(\underline{x_1,x_2}) \\
\underline{22}_f&=\partial_1\partial_2F_{-}(\underline{x_1,x_2}) \\
\underline{22}_t&=V_{--}(\underline{x_1,x_2})\label{e:vec-bilocal-def}
\end{align}
The motivation behind these definitions can be understood by appealing to what these expressions
look like in free field theory. There, they will be given by simple products of free
fields:
\begin{align}
B(x_1,x_2)&\sim :\phi(x_1)\phi^*(x_2):+:\phi(x_2)\phi^*(x_1):
\\
F_{-}(x_1,x_2)&\sim :\bar{\psi}(x_1)\gamma_-\psi(x_2):-:\bar{\psi}(x_2)\gamma_-\psi(x_1):
\\
V_{--}&\sim :F_{-\{\alpha\}}(x_1)F_{-\{\alpha\}}(x_2):
\end{align}
It is clear from the basic properties of our lightcone limits that when they are inserted into
correlation functions with another conserved current $j_s$, they will be proportional to an
appropriate free field correlator. Since $\vev{\underline{22}s} = 0$ for odd $s$, only the
correlation functions with even $s$ will be nonzero:
\begin{align}
\vev{B(\underline{x_1,x_2})j_s} &\propto
\vev{\phi(x_1)\phi^*(x_2)j_s(x_3)}_{\text{free}}\label{e:scalar-bilocal}\\
\vev{F_-(\underline{x_1,x_2})j_s} &\propto
\vev{\psi(x_1)\gamma_-\bar{\psi}(x_2)j_s(x_3)}_{\text{free}}\label{e:ferm-bilocal}\\
\vev{V_{--}(\underline{x_1,x_2})j_s} &\propto 
\vev{F_{-\{\alpha\}}(x_1)F_{-\{\alpha\}}(x_2)j_s(x_3)}_{\text{free}}\label{e:vec-bilocal}
\end{align}
Of course, away from the lightcone, things will not be so simple: we have not even defined the 
quasi-bilocal operators there, and their behavior there is the reason why they are not true
bilocals. In fact, even on the lightcone, these expressions are not fully conformally invariant:
the contractions of indices performed in equations \ref{e:ferm-bilocal} and \ref{e:vec-bilocal} are
only invariant under the action of the collinear subgroup of the conformal group defined by the line
connecting $x_1$ and $x_2$. For now, however, the lightcone properties enumerated above are enough to
establish the commutator of $Q_s$ with the bilocals. As usual, we begin with the bosonic case:

Assume that $\vev{\underline{22}_b2} \neq 0$. Our goal is to show that 
\begin{equation} 
[Q_s, B(\underline{x_1,x_2})] = (\partial^{s-1}_1+\partial^{s-1}_2) B(\underline{x_1,x_2}). 
\label{e:boson-bilocal-commutator}
\end{equation}
This can be shown using the same arguments as \cite{Maldacena:2011jn}. To begin, notice that the
action of $Q_s$ commutes with the lightcone limit. Thus,
\begin{equation}
\vev{[Q_s,B]j_k} = \vev{\underline{[Q_s,j_2]j_2}j_k} + \vev{\underline{j_2[Q,j_2]}j_k} 
= -\vev{\underline{j_2j_2}[Q_s,j_k]} = \vev{[Q_s,\underline{j_2j_2}]j_k}
\end{equation}
This immediately leads to:
\begin{equation}
[Q_s,B(\underline{x_1,x_2})] = (\partial^{s-1}_1+\partial^{s-1}_2) \tilde B(\underline{x_1,x_2}) 
+ (\partial^{s-1}_1-\partial^{s-1}_2) B'(\underline{x_1,x_2}),
\end{equation}
Here, $\tilde B$ is built from even currents, while $B'$ is built from odd
currents. This makes the whole expression symmetric. We would like to show that $B'=0$. Therefore,
suppose otherwise so that some current $j_{s'}$ has nontrivial overlap with $B'$. Then, the charge
conservation identity $0 = \vev{\left[Q_{s'},B' j_2\right]}$ yields
\begin{align}
0&=\vev{\left[ Q_{s'},B'(\underline{x_1,x_2})\right] j_2}+\vev{B'(\underline{x_1,x_2})\left[ Q_{s'},j_2\right]},
\\
\Rightarrow 0 &=\gamma \left(\partial^{s'-1}_1-\partial^{s'-1}_2\right) \vev{\underline{\phi \bar\phi} j_2} +\sum \limits_{k=0}^{s'+1}\tilde \alpha _k \partial^{s'+1-k}\vev{\underline{\phi\bar\phi}j_k}.
\end{align}
Using the same techniques as the previous section, we obtain 
\begin{equation}
0=\gamma ((p_1^+)^{s'-1}-(p_2^+)^{s'-1}) F_2(p_1^+, p_2^+)
+\sum_{k=0}^{s'+1}\tilde\alpha_{k} (p_1^++p_2^+)^{s'+1-k}F_{k}(p_1^+,p_2^+).
\end{equation}
In this sum, $\tilde\alpha_{s'}\neq0$ because $j_{s'} \subset \left[ Q_{s'},2\right]$.
Therefore, we can use the same procedure as before to show that all $\tilde\alpha_k$ are nonzero
if they are nonzero for the free field theory. In particular, since $\tilde{\alpha}_1$ is not 
zero for the complex free boson, the overlap between $j_1$ and $B'$ is not zero. Now, let's
consider 
\begin{equation}
0=\vev{\left[Q_s,Bj_1\right]}= \left( \partial_1^{s-1}-\partial_2^{s-1}\right)\vev{B'j_1}+\vev{B\left[Q_s,j_1\right]},
\end{equation}
where $Q_s$ is a charge corresponding to any even higher-spin current appearing in the operator product
expansion of $\underline{j_2j_2}_b$. We have shown the first term is not zero. We will prove that
the second term must be equal to zero to get a contradiction. Specifically, we will show that there
are no even currents in $[Q_s,j_1]$. Since $B$ is proportional to $\underline{22}$, and since
$\vev{22s} = 0$ for all odd s, this yields the desired conclusion.

Consider the action of $Q_s$ on $\vev{221}$. We obtain the now-familiar form:
\begin{equation}
0=\gamma ((p_1^+)^{s-1}-(p_2^+)^{s-1})F_1(p_1^+,p_2^+) 
+\sum_{k=0}^s\tilde\alpha_{k} (p_1^++p_2^+)^{s-k}F_{k}(p_1^+,p_2^+)
\label{e:221-identity}
\end{equation}
We want to show that $\alpha_k = 0$ for even $k$. Recall the definition of $F_k$:
\begin{align}
F_k &= (p_2^+)^k{_2}F_1\left(2-\frac{d}{2}-k,-k,\frac{d}{2}-1,-\frac{p_1^+}{p_2^+}\right)\\
&= \sum_{i=0}^k c^k_i (p_1^+)^i(p_2^+)^{s-i}
\end{align}
The hypergeometric coefficients $c^k_i$ have the property that $c^k_i = (-1)^kc^k_{k-i}$. 
Now, we collect terms in equation (\ref{e:221-identity}) proportional to
$(p_1^+)^s$ and $(p_2^+)^s$ - each sum must vanish separately for the entire polynomial to
vanish. We obtain
\begin{align}
\gamma + \sum_{0 \le k \le s \text{ odd}}\alpha_ku_k + \sum_{0 \le k \le s \text{ even}}\alpha_kv_k 
= 0 \\
- \gamma - \sum_{0 \le k \le s \text{ odd}}\alpha_ku_k + \sum_{0 \le k \le s \text{ even}}\alpha_kv_k
= 0
\end{align}
Here, $u_k$ and $v_k$ are sums of products of coefficients of the hypergeometric function
and the binomial expansion of $(p_1^++p_2^+)^{s-k}$; we do not care about their
properties except that, with the signs indicated above, they are strictly positive, as can be
verified by direct calculation. By adding and subtracting these equations, we obtain two
separate equations that must be satisfied by the odd and even coefficients separately
\begin{align}
\gamma + \sum_{0 \le k \le s \text{ odd}}\alpha_ku_k = 0 \\
\sum_{0 \le k \le s \text{ even}}\alpha_kv_k = 0
\end{align}
Exactly analogously, we may do the same procedure to every other pair of monomials
$(p+1^+)^a(p_2^+)^{s-a}$ and $(p_1^+)^{s-a}(p_2^+)^a$ to turn the constraints
for the two monomials into constraints for the even and odd coefficients (where we're considering
$\gamma$ as an odd coefficient) separately. Hence, by multiplying each term by the monomial from
which it was computed and then resumming, we find that the original identity (\ref{e:221-identity})
actually splits into two separate identities that must be satisfied. For the even terms, this
identity is:
\begin{equation}
0 = \sum_{0 \le k \le s \text{ even}} \alpha_k(p_1^++p_2^+)^{s-k} (p_2^+)^k
{_2}F_1\left(2-\frac{d}{2}-k,-k,\frac{d}{2}-1,-\frac{p_1^+}{p_2^+}\right)
\end{equation}
Then, we may again use the argument from section \ref{charge-conservation-identities} to conclude
that all $\alpha_k = 0$ for even $k$, which is what we wanted. Thus, $B'=0$.

Now we would like to show that $B=\tilde B$. First of all we will show that $\tilde B$ is nonzero.
Consider the charge conservation identity 
\begin{equation}
0 = \vev{\left[Q_s, B j_2\right]} = \left(\pd^{s-1}_1 + \pd^{s-1}_2\right)\vev{\tilde{B}2} +
\vev{B,[Q_s,2]}
\end{equation}
Since $\left[Q_s, j_2\right]\supset\partial j_s$, and since $\vev{Bs} \neq 0$, the second term in
that identity is nonzero, and so $\tilde B$ must be nonzero. Now we can
normalize the currents in such a way that $j_2$ has the same overlap with $\tilde B$ and $B$. After
normalization, we know that $B-\tilde B$ does not contain any spin $2$ current because the stress
tensor is unique, by hypothesis. Now, we will show that $B-\tilde B$ is zero by contradiction.
Suppose $B-\tilde B$ is nonzero. Then, there is a current $j_s$ whose overlap with $B-\tilde B$ is 
nonzero. Then, the charge conservation identity for the case $s>2$
is 
\begin{align}
0&=\vev{\left[Q_s,\left( B-\tilde B\right) j_2\right]},\\
0&=\gamma \left( \left(p_1^+\right)^{s-1}+\left(p_2^+\right)^{s-1}\right)F_2(p_1^+,p_2^+)+ 
\sum\limits_{k=0}^{s+1}\tilde \alpha_k (p_1^++p_2^+)^{s+1-k} F_{k}\left(p_1^+,p_2^+\right),
\end{align}
where we assume that $\tilde\alpha_s\neq 0$. Then, we can again run the same analysis as section
\ref{charge-conservation-identities} to conclude
that since $\tilde \alpha_s\neq 0$, we must have $\tilde \alpha_2\neq 0$ - that is, $j_2$ has
nonzero overlap with $B-\tilde B$, which is a contradiction. It means that $B-\tilde B$ has no
overlap with any currents $j_s$ for $s>2$. The only possibility is to overlap only with spin zero
currents. Suppose that there is a current $j'_0$ that overlaps with $B-\tilde{B}$, where the prime
distinguishes it from a spin $0$ current $j_0$ that could appear in $B$. We first show that
$\vev{j_0j_0'} = 0$. Consider the charge conservation identity the action $Q_4$ on $\vev{(B-\tilde B)j_0}$.
The action of the charge is $\left[Q_4,0\right]=\partial^3 j_0+\partial j_2+\dots$, where the
$\dots$ represent terms that cannot overlap with $\underline{22}$ (from which $B$ is constructed) or
the even currents that appear in $\tilde{B}$. By hypothesis, $B-\tilde B$ has no overlap with $j_2$,
so the identity simplifies to $\vev{j_0j_0'} = 0$. Then, since $j_0'$ is nonzero, it should have
nontrivial overlap with some $Q_s$. Now, recall the fact that if a current $j'$ appears (possibly
with some number of derivatives) in the commutator of $[Q_s,j]$, then $j$ appears in $[Q_s,j']$.
Thus, there should be a current current of spin $s''<s$ such that $\left
[Q_s,j_{s''}\right]=j_0'+\dots$. The action $Q_s$ on $\vev{\left(B-\tilde B \right)j_{s''}}$ is 

\begin{equation}
 \vev{\left[ Q_s,\left(B-\tilde B \right)j_{s''}\right]}=\partial^3_3\vev{\left(B-\tilde B \right)j_0'} +\partial\vev{\left(B-\tilde B\right) j_2},
\end{equation}
Here, we have used that the action of $Q_s$ on $B$ and $\tilde{B}$ is identical because $B' = 0$.
Then, since the second term is zero, thus the first term is equal to zero as well. Thus, $B-\tilde
B$ has no overlap with any currents and is equal to zero, as desired.

In the fermionic case, we can run almost the same argument as in the bosonic case, except there is no
discussion of a possible $j_0$, since there is no conserved spin zero current in the free fermion
theory. We obtain the action of the charge on the fermionic quasi-bilocal is 
\begin{equation}
 [Q_s, F_{-}(\underline{x_1,x_2})] = (\partial^{s-1}_1+\partial^{s-1}_2) F_{-}(\underline{x_1,x_2}). 
\end{equation}

In the tensor case, we again can repeat the argument to obtain
\begin{equation}
 [Q_s, V_{--}(\underline{x_1,x_2})] = (\partial^{s-1}_1+\partial^{s-1}_2) V_{--}(\underline{x_1,x_2})
\end{equation}
In this case, there is neither a conserved spin $0$ or spin $1$ current in the free tensor theory.
The argument works the same way, however, if we consider $j_3$ instead of $j_1$ in the steps of the
argument that require it.

\section{Quasi-bilocal fields: correlation functions}\label{qb-correlators}
In this section, we will discuss how to precisely define the quasi-bilocal operators in a way that
makes their symmetries manifest. 
In particular, each of the three bilocals will be bi-primary operators in some sense. 
This will allow us to argue that the correlation functions of the bilocals
should agree with an appropriate corresponding free-field result. We will then explore what this
implies for the full theory in section \ref{constraining-correlators}. 
\subsection{Symmetries of the quasi-bilocal operators}
Let us first consider the case of the bosonic bilocal operator $B(x_1,x_2)$. Recall 
that, on the lightcone, the bilocals should imitate products of the appropriate free fields. In the
bosonic free-field theory, the operator product expansion of $\phi(x_1)\phi^*(x_2)$ is composed
of all of the even-spin currents of the theory with appropriate numbers of derivatives and factors
of $(x_1-x_2)$ so that the expression has the correct conformal dimension. More explicitly, we may
write:
\begin{align}
\phi(x_1)\phi^*(x_2) &= \sum_{\text{even } s\ge 0} b^{\text{free}}_s(x_1,x_2) \\
b^{\text{free}}_s(x_1,x_2) &= \sum_{(k,l) \mid s+l-k = 0} c_{skl}(x_1-x_2)^k\pd^lj_s\left(\frac{x_1+x_2}{2}\right)
\label{e:free-decomposition}
\end{align}
All the coefficients $c_{skl}$ may be computed exactly in the free theory just by Taylor expansion.
We have shown that all the currents $j_s$ exist in our theory for all even $s$. So we may define 
an analogous quantity in our theory as follows:
\begin{align}
B(x_1,x_2) &= \sum_{\text{even } s\ge 0} b_s(x_1,x_2) \\
b_s(x_1,x_2) &= \sum_{(k,l) \mid s+l-k=0} c'_{skl}(x_1-x_2)^k\pd^lj_s\left(\frac{x_1+x_2}{2}\right)
\label{e:decomposition}
\end{align}
Here, the $c'$ are some other coefficients which are to be determined by demanding that this
definition of $B$ coincide with the definition given on the lightcone in the previous section,
i.e.~that $\partial_1^2\partial_2^2B(\underline{x_1,x_2}) = \underline{22}_b$. We claim that this
can be accomplished by choosing the $c'$ coefficients such that 
$\vev{B(\underline{x_1,x_2})j_s} \propto \vev{\underline{\phi(x_1)\phi^*(x_2)}j_s}_{\text{free}}$. 
To see that there exists such a choice of $c'$ which can achieve this condition, we explicitly
compare $\vev{Bj_s}$ and $\vev{\phi\phi^*j_s}_\text{free}$ term by term using
\ref{e:free-decomposition} and \ref{e:decomposition}. Each term in both of
these correlation functions has the structure $(x_1-x_2)^k\partial^l\vev{j_{s'}j_s}$ with
coefficient $c_{s'kl}$ and $c'_{s'kl}$, respectively. Two-point functions of primary operators in
CFT's are determined up to a constant, so each term is identical up to a possible scaling, which can
be eliminated by choosing the $c'$ coefficient appropriately. Then, by applying
$\partial_1^2\partial_2^2$ to both sides of 
$\vev{B(\underline{x_1,x_2})j_s} \propto \vev{\underline{\phi(x_1)\phi^*(x_2)}j_s}_{\text{free}}$,
we find that our definition coincides on the lightcone, as desired. This
construction works the same way for the fermionic and tensor quasi-bilocals with analogous results,
except that the quasi-bilocals in those cases carry the appropriate spin structure.

Since the conformal transformation properties of a conserved current $j_s$ is theory-independent in
the sense that it is completely fixed by its spin and conformal dimension, it is manifest from this
definition that the bosonic quasi-bilocal $B(\underline{x_1, x_2})$ has the same transformation
properties under the full conformal group as a product of free bosons. That is, it is a scalar
bi-primary field with a conformal dimension of $1$ with respect to each argument.

On the other hand, consider the fermonic and tensor quasi-bilocals $F_-$ and $V_{--}$. The same
line of reasoning tells us that they will transform like products of free fields contracted in a
particular way: $F_-$ will transform like $:\psi \gamma_- \bar{\psi}:$ does in the free fermionic
theory, and $V_{--}$ will transform like $:F_{-\{\alpha\}} F_-^{\{\alpha\}}:$ does in the theory of
a free $\frac{d-2}{2}$-form\footnote{Technically, the argument given above for the symmetries of the 
bosonic quasi-bilocal only works for even dimensions in the tensorial case since the free
$\frac{d-2}{2}$-form exists only in even dimensions, so the matching procedure can't be carried out
naively in odd dimensions. On the other hand, it is evident from the definition
\ref{e:vec-bilocal-def} that it has at least the collinear conformal symmetry since there are no
derivatives to be ``integrated out.''}. 
These contractions, however, are not preserved by the full conformal
group - the special conformal transformations orthogonal to the $-$ direction will ruin the
structure of the Lorentz contraction. Thus, even in the free theory, these objects are not
preserved by the full conformal group. They are only preserved by the so-called
collinear conformal group generated by $K_-, P_+, J_{+-},$ and $D$, where $K, P, J,$ and $D$ are the
generators of special conformal transformations, translations, boosts, and dilatations,
respectively. It is clear from the structure of the conformal algebra that the commutation
relations of this subset of conformal generators closes, so it forms a proper sub-algebra. Thus,
what we are allowed to conclude is that $F_-$ and $V_{--}$ are bi-primary operators with respect to
this collinear subgroup, not the conformal group. Nevertheless, this will still be enough symmetry
for our purposes. 

The key fact which is still true for this more restricted set of symmetries is that under $K_-$, the
special conformal transformation in the $-$ direction, the $n$-point function of fermionic and
tensor quasi bi-primaries should scale separately in each variable. That is, under $K_-$, if
$x \rightarrow x'$ and $g_{\mu\nu}(x) \rightarrow \Omega^2(x) g_{\mu\nu}(x)$, we have 
\begin{multline}
\vev{F_{-}(\underline{x_1',x_2'}),\cdots,F_{-}(\underline{x_{2n-1}',x_{2n}'})} \\
= \Omega(x_1)^{d/2-1}\dots\Omega(x_{2n})^{d/2-1}\vev{F_{-}(\underline{x_1,x_2}),\cdots,F_{-}(\underline{x_{2n-1},x_{2n}})}
\label{e:fermion-primary}
\end{multline}
and
\begin{multline}
\vev{V_{--}(\underline{x_1',x_2'}),\cdots,V_{--}(\underline{x_{2n-1}',x_{2n}'})} \\
= \Omega(x_1)^{d/2-1}\dots\Omega(x_n)^{d/2-1}\vev{V_{--}(\underline{x_1,x_2}),\cdots,V_{--}(\underline{x_{2n-1},x_{2n}})}
\label{e:vector-primary}
\end{multline}
The proof of these two statements is given in appendix \ref{primary-proof}.
\subsection{Correlation functions of the bosonic quasi-bilocal}
Now we will discuss the structure of the $n$-point functions of the quasi-bilocals. Again, let's
begin with the bosonic case. We wish to constrain $\vev{B(\underline{x_1,x_2}) \dots
B(\underline{x_{2n-1},x_{2n}})}$. We established that $B(x_1, x_2)$ has the transformation
properties of a product of two free fields under the full conformal group - i.e.~ it is a bi-primary
field with dimension $\frac{d-2}{2}$ in each variable. That means that the
$n$-point function can only depend on distances between coordinates $d_{ij}$ and have conformal
dimension $\frac{d-2}{2}$ with respect to each variable. Since $x_1$ and $x_2$ are lightlike separated,
$d_{12}$ cannot appear, and similarly for every pair of arguments of the same bilocal. There is
also a permutation symmetry: $B$ is symmetric in its two arguments, and the $n$ point function must
be symmetric under interchange of any pair of the identical $B$'s. Finally, there is the
higher-spin symmetry. In the bosonic case, the charge conservation identity
(\ref{e:boson-bilocal-commutator}) imposes the simple relation
\begin{equation}
\sum_{i=1}^{2n} \pd^{s-1}_i \vev{B(\underline{x_1,x_2})\dots B(\underline{x_{2n-1}x_{2n}})}, 
\,\,\,\,\,\, \text{for all even s} 
\label{e:important-cci}
\end{equation}
As shown in appendix E of \cite{Maldacena:2011jn}, this fixes the $x^-$ dependence of the 
$n$-point function to have the particular form: 
\begin{equation}
\sum_{\sigma \in S^{2n}} g_{\sigma}\left(x^-_{\sigma(1)} - x^-_{\sigma(2)}, x^-_{\sigma(3)} -
x^-_{\sigma(4)},\dots, x^-_{\sigma(2n-1)}-x^-_{\sigma(2n)}\right) 
\label{e:pairwise-differences}
\end{equation}
where $S^{2n}$ is the set of permutations of $2n$ elements. The point is that the $x_i^-$
dependence of the $n$-point function is constrained such that, for each $g_\sigma$, $x_i^-$ 
can only appear in a difference with one and only one other coordinate. This is
a very strong constraint. The conformal symmetry tells us that each $g_\sigma$ in the above series
can be written as a product of a dimensionful function of distances with the correct dimension in
each variable times a smooth, dimensionless function of conformal cross-ratios. The constraint on
the functional form of $g_\sigma$, however, forbids all such functions except the trivial function
$1$, because each cross ratio separately violates the constraint.

Putting it all together, we conclude that the $n$-point function has to be proportional to a sum of
terms with equal coefficients, each of which is a product $\prod d_{ij}^{-(d-2)}$, where the product has
$n$ terms corresponding to some partition of the $2n$ points into pairs where no pair contains two
arguments of the same bilocal. For example, the two-point function is:
\begin{equation}
\vev{B(\underline{x_1,x_2})B(\underline{x_3,x_4})} = \tilde{N}_b\left(\frac{1}{d_{13}^{d-2}d_{24}^{d-2}} 
+ \frac{1}{d_{14}^{d-2}d_{23}^{d-2}}\right),
\end{equation}
where $\tilde{N}_b$ is a constant of proportionality. One immediately notes that the expressions one
obtains this way for all $n$-point functions of the quasi-bilocals are proportional to the $n$-point
function of $:\phi(x_1)\phi(x_2):$ in a theory of free bosons.
\subsection{Correlation functions of the fermionic and tensor quasi-bilocal}
In the fermionic and tensor cases, we claim that the correlation functions of the quasi-bilocals still
coincide with the correlation functions of the corresponding free field theories, despite the fact
that the fermionic and tensor quasi-bilocals have less symmetry than the bosonic quasi-bilocal. The
argument, however, is somewhat more complicated due to the reduced amount of symmetry. The proof is
essentially the same for both the fermionic and tensor cases, so we will only present the
argument for the tensor case. Our general strategy will be to compare the constraints that one
obtains from the definition of $V_{--}$ as the lightcone limit $\underline{22}_t$ with the constraints
one obtains from the symmetries of $V_{--}$ as established by its definition away from the lightcone
given at the beginning of this section. In the bosonic case, we only used the latter, but in the
fermionic case and tensor case, we will need the former as well.

First, we consider what the $2n$-point function of $T_{--}$ is away from the lightcone. We know
from the definition of $V_{--} = \underline{22}_t$ that if we take $n$ lightcone limits of this $2n$
point function in each pair of adjacent arguments $(x_1, x_2), (x_3, x_4), \dots (x_{2n-1},x_{2n})$,
we will obtain the $n$ point function of $V_{--}(\underline{x_1,x_2})$. It may be the case that the
definition of $V_{--}$ given earlier as a sum of currents and descendants (with appropriate
derivatives and powers of $x$) will yield a different result away from the lightcone, but
nevertheless, it must agree with the $2n$-point function of $T_{--}$ in the lightcone limit.

Generically, the $2n$ point function of $T_{--}$ with arguments in arbitrary locations can be
decomposed as a polynomial in some basis of conformally invariant structures. One convenient basis
is the $\{H_{ij},V_i\}$ space defined in \cite{Costa:2011mg}. In this basis, we may write
\begin{equation}
\vev{T_{--}(x_1) \cdots T_{--}(x_{2n})} = \frac{\vev{\vev{T_{--}(x_1) \cdots T_{--}(x_{2n})}}}
{d_{12}^{d-2}d_{23}^{d-2}\dots d_{2n-1,2n}^{d-2}d_{2n,1}^{d-2}}
\label{e:double-vev}
\end{equation}
where
\begin{equation}
\vev{\vev{T_{--}(x_1) \cdots T_{--}(x_{2n})}} 
= \sum_i f_i(\{u_j\}) \left(\prod_{k<l} H_{kl}^{h_i^{(k,l)}}\right) \left(\prod_{k < l < m}
V_{k,lm}^{v_i^{(k,lm)}}\right)
\end{equation}
where $f_i(\{u_j\})$ is an arbitrary function of cross-ratios $\{u_j\}$, the $h_{kl}$ and $v_i$
coefficients satisfy 
\begin{equation}
\sum_{l,m | k < l < m} v_i^{(k,lm)} + \sum_{n | k < n} h_i^{(k,n)} = 2 \,\,\text{ for all } i, k
\end{equation}
and the conformal invariants are
\begin{align}
V_{k, lm} &= \frac{x_{kl}^+}{d_{kl}^2} + \frac{x_{km}^+}{d_{km}^2} \\
H_{kl} &= \frac{-2(x_{kl}^+)^2}{d_{kl}^4}
\end{align}
Note that this decomposition omits structures which contain the epsilon tensor, which all vanish
in our formalism because we contract all free indices with the same polarization vector in the $-$
direction. 

We would like to understand the properties of this decomposition under the tensor lightcone limit
\ref{e:vector-lcl}. First, note that the universal dimensionful factor of distances that is
factored out of $\vev{\vev{T\dots T}}$ in \ref{e:double-vev} is conventional. In principle, one
could choose it to be something different and compensate by appropriate redefinitions of $f_i$. We
have chosen it as shown in order to simplify the structure of this function under the lightcone
limit. More precisely, the distances corresponding to pairs of points that become lightlike
separated $d_{12}$, $d_{34}$, \dots, $d_{2n-1,2n}$ vanish in the lightcone limit, so they cannot
explicitly appear in the correlation function, and we have chosen the universal factor so that this
property is manifest. To see this, note that when we take the lightcone limit \ref{e:vector-lcl}
of this general structure, the part of this universal factor corresponding to the distances between
points that become lightlike separated - i.e.~$d_{12}^{-d+2}d_{34}^{-d+2}\dots d_{2n-1,2n}^{-d+2}$ -
becomes $d_{12}^4d_{34}^4\dots d_{2n-1,2n}^4$. This residual factor is exactly cancelled out by the
$V$ and $H$ terms corresponding to the $x^+$ factors stripped away in \ref{e:vector-lcl}. To see
this, recall that the light-cone limits of correlation functions are well-defined and
non-divergent\footnote{As we remarked before, this is only true a priori if we subtract off the
bosonic and fermionic pieces, but we will show in section \ref{constraining-correlators} that if any
one of the three lightcone limits are nonzero, it follows that the other two are zero, so this
subtraction procedure is not actually necessary.}, so any structure consistent with conformal
symmetry needs to appear with enough $V$'s and $H$'s with appropriate indices to cancel out the
factors of $(x_{12}^+)^{-2}, (x_{34}^+)^{-2}, \dots, (x_{2n-1,2n}^+)^{-2}$ that appear in the
lightcone limit. As noted earlier, these factors of $V$'s and $H$'s come with exactly two powers
each of $d_{12}^2, \dots, d_{2n-1,2n}^2$, which is exactly what is needed to cancel out the residual
term.

Thus, after the lightcone limit, the most general structure that can appear in the $n$-point
function of $V_{--}$ is:
\begin{align}
\vev{V_{--}(\underline{x_1,x_2}) \dots V_{--}(\underline{x_{2n-1},x_{2n}})}
&= \vev{\underline{T_{--}(x_1)T_{--}(x_2)}_t \cdots \underline{T_{--}(x_{2n-1})T_{--}(x_{2n})}_t}\\
&= \sum_i \frac{f_i(\{u_j\})}{d_{23}^{d-2}\dots d_{2n,1}^{d-2}}
\prod_{k,l} \left(\frac{x_{kl}^+}{d_{kl}^2}\right)^{c_{kl}}
\label{e:general-structure}
\end{align}
where the product over $k$ and $l$ is understood to be restricted to pairs $(k,l)$ not corresponding
to $x_k, x_l$ lightlike separated, and $\sum c_{kl} = 2n$.

We can determine which terms of this form are consistent with the symmetries of $V_{--}$.
Consider the $n$-point correlation function of $V_{--}$. Its transformation properties under 
Lorentz transformations and dilatations tell us that we must have $2n$ $+$ indices in the
numerator of the correlation function, and that the overall scaling dimension of the $n$-point
function should be $2n \times d/2 = dn$. Then, as mentioned before, since $V_{--}$ is a bi-primary
under the collinear conformal group, the $n$-point function should scale appropriately in each
variable separately after acting with $K_-$ according to \ref{e:vector-primary}. 
In order to satisfy this constraint, for each independent structure appearing in the correlation
function and each index $i$, we must have $2$ factors of $x_{ij}^+$ in the numerator (not necessarily
the same $j$ for each of the $2$ factors) and $d+2$ powers of $d_{ik}$ in the denominator for some
$k$ (again, not necessarily the same $k$ for each of the $d+2$ factors). Once we have picked such a
denominator, there is still some ambiguity since conformally invariant functions $f_i$ can still 
appear after imposing this constraint (since they are fixed by $K_-$), and such functions can change
the denominator. What is tightly constrained here is the numerator - i.e.~the spin structure. 
``Imbalanced'' structures with that would otherwise be allowed by Lorentz symmetry, scaling
symmetry, and permutation symmetry cannot appear. For example, for the two-point function
$\vev{V_{--}V_{--}}$ in four dimensions, structures such as $\frac{(x_{13}^+)^4}{d_{13}^{12}} +
\frac{(x_{24}^+)^4}{d_{24}^{12}}$ do not satisfy \ref{e:vector-primary}. Note that the numerators
which are allowed by this constraint are precisely the ones that appear in free-field correlation
functions (i.e.~the ones arising from Wick contractions of free fields) and no others.

Now, let's impose the higher-spin constraint, which stipulates that the correlation function 
must be a sum of terms $g_\sigma$ which have the functional 
form given by $\ref{e:pairwise-differences}$. Since that constraint only involves the dependence in
the $x^-$ direction, it does not constrain the numerator, which involves only terms involving the 
$x_i^+$ variables. However, it does restrict the denominator to only have each index $i$ involved in a
power $d_{ik}$ for one specific $k$ since $d_{ik}$ does depend nontrivially on $x_{ik}^-$. That is,
the denominator is built out of terms like $d_{ik}^{d+2}$. This constrains the $f_i$ powerfully. 
Since each cross ratio separately violates the higher spin constraint, the only $f_i$ that
can appear are the ones whose product with a denominator satisfying the higher-spin constraint
is another denominator satisfying that constraint. That is, once we have picked a denominator, 
the $f_i$ can only be very specific kinds of rational functions. We can still generate terms that
don't appear in the free-field result, however, because
the spin structure in the numerator doesn't have to match the index structure of the denominator.
For example, the following structure could in principle appear in the four-point function of
$V_{--}$, but obviously this structure is not generated in the free theory:
\begin{equation}
\frac{(x_{14}^+)^2 (x_{27}^+)^2 (x_{36}^+)^2 (x_{58}^+)^2}
{(d_{13} d_{24} d_{57} d_{68})^{d+2}}
\label{e:nice-structure}
\end{equation}
This structure has a numerator which is consistent with free field theory but a denominator that
does not match the result one would obtain from the free field propagator. Another possibility is
to write a structure where the numerator corresponds to the connected part of the free-field
correlator - i.e.~the two factors of $x_{ij}^+$ appear with different $j$ for some $i$.
\begin{equation}
\frac{x_{13}^+ x_{32}^+ x_{28}^+ x_{86}^+ x_{67}^+ x_{75}^+ x_{54}^+ x_{41}^+}
{(d_{13} d_{24} d_{57} d_{68})^{d+2}}
\label{e:bad-structure}
\end{equation}
Purely on symmetry considerations, these terms are consistent with the general structure
\ref{e:general-structure}. Indeed, one can set \ref{e:nice-structure} and \ref{e:bad-structure}
equal to \ref{e:general-structure} to explicitly solve for the function $f_i(\{u_j\})$ that
generates it, and one can check that this $f_i$ is indeed conformally invariant, as required. 
These structures are inconsistent, however, with cluster decomposition. To see this, we
examine the tensor analogue of \ref{e:decomposition}:
\begin{align}
V_{--}(x_1,x_2) &= \sum_{\text{even } s\ge 2} v^s_{--}(x_1,x_2) \\
v^s_{--}(x_1,x_2) &= \sum_{k,l}
c_{kl}(x_1-x_2)^k\pd^lj_s\left(\frac{x_1+x_2}{2}\right)\label{e:tensor-decomposition}
\end{align}
Comparing the conformal dimension of the left and right hand side yields the constraint that 
$s+l-k = 2$. Hence, by setting $x_1=x_2$, we extract the $k=0$ piece, forcing $l = 0$ and 
$s = 2$ (since $s=1$ is not realized in the tensor sector). That is, 
$V_{--}(x,x) = T_{--}(x)$. By performing this projection on each factor of $V_{--}$ in the
correlation function (i.e.~setting $x_1 = x_2$, $x_3 = x_4$, etc.), we obtain an expression for the
$n$-point function of $T_{--}$, which we know must satisfy cluster decomposition since $T$ is a
local operator. Then, by taking
the points to be separated very far apart from each other, we obtain constraints on how the
structures must simplify. For example, we know that if we take $x_1$ and $x_3$ to be very far from
all the other points, we must have that
\begin{equation}
\vev{T_{--}(x_1) T_{--}(x_3) \dots T_{--}(x_{2n-1})} \Longrightarrow 
\vev{T_{--}(x_1)T_{--}(x_3)}\vev{T_{--}(x_5) \dots T_{--}(x_{2n-1})}
\end{equation}
This factorization property is not satisfied by the structure \ref{e:nice-structure}, for example.
Indeed, the only way to satisfy all such constraints arising from cluster decomposition is to have
all powers of $x_{ij}^+$ appear with the corresponding factor of $d_{ij}^{d-2}$ in the denominator,
modulo trivial equalities such as $x_{13}^+ = x_{14}^+$ (which arise since points which are taken to
be $-$ separated in the lightcone limit have the same difference in the $+$ direction). If it
appears with the wrong $d_{ij}$ factor in the denominator (again, modulo the trivial relabelings of
the spin structure), it cannot satisfy the cluster decomposition identity arising from taking the
two points appearing in that factor to be very far from all the other points. The spin structure
required by the factorization will simply not be present.

Hence, the only allowed terms are the ones that are built from free-field propagators 
$(x_{ij}^+)^2/d_{ij}^{d+2}$. Permutation symmetry implies that the coefficients of all the
structures that can appear are the same up to disconnected terms which are fixed, as before, by
cluster decomposition. This implies that the $n$-point function of bilocals $V_{--}$ are exactly
the same as the $n$-point function of stress tensors in free field theory up to a possible overall
constant. 

Clearly, this entire argument works for the fermionic case as well with only minor modifications - 
the projection procedure that isolates the contribution from the stress tensor is slightly more
complicated since it appears at first order, not zeroth order, in $x_{12}$ in the fermionic 
analogue of \ref{e:tensor-decomposition}, and the correlation function is permutation anti-symmetric
instead of symmetric because fermions anticommute. All other steps are the same, and we conclude
that in the fermionic case, the $n$-point functions of bilocals are also given by the free field
result. For example, the two-point functions of fermionic and tensor quasi-bilocals are given by
\begin{align}
\vev{F_-(\underline{x_1,x_2})F_-(\underline{x_3,x_4})} &=
\tilde{N}_f\left(\frac{x_{13}^+x_{24}^+}{d_{13}^{d}d_{24}^{d} }-
\frac{x_{14}^+x_{23}^+}{d_{14}^{d}d_{23}^{d}}\right)\\
\vev{V_{--}(\underline{x_1,x_2})V_{--}(\underline{x_3,x_4})} &=
\tilde{N}_t\left(\frac{(x_{13}^+)^2(x_{24}^+)^2}{d_{13}^{d+2}d_{24}^{d+2}} +
\frac{(x_{14}^+)^2(x_{23}^+)^2}{d_{14}^{d+2}d_{23}^{d+2}}\right)
\end{align}
where $\tilde{N}_f$ and $\tilde{N}_t$ are overall constants that we will presently analyze.
\subsection{Normalization of the quasi-bilocal correlation functions}
Now, let's fix the the overall constants $\tilde{N}_b$, $\tilde{N}_f$, and $\tilde{N}_t$ in front of
each $n$-point function. We claim that they all are fixed by the normalization of the two-point
function of the bilocals. This can be seen by considering how one can obtain the $n$-point function
of quasi-bilocals from the $n-1$ point function. We know the $n$-point function of some
quasi-bilocal $\mathcal{A}$ is:
\begin{equation}
\underbrace{\vev{\mathcal{A}\dots\mathcal{A}}}_{n \text{ copies of } \mathcal{A}} = \tilde{N}_ng(d_{ij}) 
\end{equation}
where $g$ is some known function which agrees with the result for the
$n$-point function of the corresponding free theory bilocal. Each bilocal contains the stress
tensor $j_2$ in its OPE, so we can consider acting on both sides with the projector $P$ which
isolates the contribution of $j_2$ from the first bilocal. We have already seen, for example, 
that for the tensor bilocal, this projector just sets $x_1 = x_2$. Then, we can integrate 
over the coordinate $x_1$. This yields the action of the
dilatation operator on the $n-1$ point function, whose eigenvalue will be some multiple of the
conformal dimension of the appropriate free field. So by this procedure, we can fix the coefficient
in front of the $n$-point function in terms of the $n-1$ point function. So by recursion, all the
coefficients of the correlation functions are fixed by the coefficient $\tilde{N}$ appearing in
front of the two-point function.

\section{Constraining all the correlation functions}\label{constraining-correlators}
We have shown now that the $n$-point functions of all the quasi-bilocal fields exactly coincide with
the corresponding free-field result for a theory of $N$ free fields of appropriate spin for some $N$
(which we will show later must be an integer). 
Now, we will explain how to use these facts to constrain all the other correlation functions of the theory. 
We will start by proving that the three point function $\vev{222}$
must be either equal to the result for a free boson, a free fermion, or a free $\frac{d-2}{2}$ form. 
That is, if we write the most general possible form:
\begin{equation}
\vev{222} = c_b\vev{222}_{\text{free boson}} + c_f\vev{222}_{\text{free fermion}} 
+ c_t\vev{222}_{\text{free tensor}},
\end{equation}
then the result will be consistent with higher-spin symmetry only if $(c_b,c_f,c_t) \propto (1,0,0)$
or $(0,1,0)$ or $(0,0,1)$. 

We first show that if $\vev{\underline{22}_b2} \neq 0$ then $\vev{\underline{22}_f 2} = 0 =
\vev{\underline{22_t}2}$. Consider the action of $Q_4$ on $\vev{\underline{22}_b2}$. By exactly the
same analysis as the charge conservation identities of section \ref{charge-conservation-identities},
we obtain exactly the same expression as equation (\ref{e:bosonic-charge-identity}), except the
summation starts from $j=0$. Thus, the existence of the spin $4$ current implies the existence of
a spin $0$ current with $\vev{\underline{22}_b0} \neq 0$. The action of charge $Q_4$ on $j_0$
is 
\begin{equation}
[Q_4,j_0]=\partial^3 j_0+\partial j_2+\mbox{no overlap with }\underline{22}_b
\end{equation} 
Now consider the charge conservation identities arising from the action of
$Q_4$ on $\vev{\underline{22}_f0}$ and $\vev{\underline{22}_t0}$. Since $\vev{\underline{22}_f0}
= 0 = \vev{\underline{22}_t0}$, we conclude $\vev{\underline{22}_f 2} = 0 = \vev{\underline{22}_t2}$, as
desired.

Now, assume that $\vev{\underline{22}_b2} = 0$. It suffices to show that if
$\vev{\underline{22}_t2} \neq 0$, then $\vev{\underline{22}_f2} = 0$. In this case, by hypothesis,
the quasi-bilocal $V_{--}$ is nonzero. The results of the previous section tell us that the three
point function of the tensor quasi-bilocal is proportional to:
\begin{equation}
\vev{V_{--}(\underline{x_1,x_2})V_{--}(\underline{x_3,x_4})V_{--}(\underline{x_5,x_6})} 
\propto\frac{\left(x_{13}^+\right)^2 \left(x_{25}^+\right)^2\left(
x_{46}^+\right)^2}{d_{13}^{\frac{d}{2}+2}d_{25}^{\frac{d}{2}+2}d_{46}^{\frac{d}{2}+2}}+perm.\label{e:vecstruc}
\end{equation}
and this precisely coincides with the three-point function of the free field operator $v_{--}(x_1,x_2) =
:F_{-\{\alpha\}}(x_1)F_{-\{\alpha\}}(x_2):$
\begin{equation}
\vev{V_{--}(x_1,x_2)V_{--}(x_3,x_4)V_{--}(x_5,x_6)} \propto \vev{v_{--}(x_1,x_2)v_{--}(x_3,x_4)v_{--}(x_5,x_6)}
\label{e:vvv}
\end{equation}
Now, take $x_1$ and $x_2$ very close together and expand both sides of this equation
in powers of $(x_1-x_2)$. The zeroth order term of $v$ is clearly the normal ordered product
$:F_{-\alpha}(\frac{x_1+x_2}{2})F_{-\alpha}(\frac{x_1+x_2}{2}):$ - this is precisely the free field
stress tensor. On the other hand, we know from the previous section that the term in $V_{--}$ which
is zeroth order in $(x_1-x_2)$ - i.e.~the term that arises from setting $x_1 = x_2$, is just the
stress tensor of the theory $T_{--}$. Repeating the same procedure for the
pairs of coordinates $(x_3,x_4)$ and $(x_5,x_6)$, we obtain the desired result:
\begin{align} 
\vev{222} &= \vev{222}_{\text{free tensor }} \\
\Rightarrow \vev{\underline{22}_f2}&=\vev{\underline{22}_b2} = 0
\end{align}
as required. Therefore, since the stress-energy tensor is unique,
\begin{align}
\vev{222}_b&\neq0 \Rightarrow &\vev{222}_f&=0,& \vev{222}_t&=0,& \underline{j_2j_2}_b=&\sum_{k=0}^{\infty} \left[j_{2k}\right],& \underline{j_2j_2}_f&=0,&\underline{j_2j_2}_t &=0,
\\
\vev{222}_f&\neq0 \Rightarrow &\vev{222}_b&=0, &\vev{222}_t&=0,& \underline{j_2j_2}_f=&\sum_{k=1}^{\infty} \left[j_{2k}\right],& \underline{j_2j_2}_b&=0,&\underline{j_2j_2}_t &=0,
\\
\vev{222}_t&\neq0 \Rightarrow &\vev{222}_b&=0,& \vev{222}_f&=0,& \underline{j_2j_2}_t=&\sum_{k=1}^{\infty} \left[j_{2k}\right],& \underline{j_2j_2}_b&=0,&\underline{j_2j_2}_f &=0,
\end{align}
where square brackets denotes currents and their descendants. This establishes the claim that the
three-point function of the stress tensor coincides with the answer for some free theory.

At this point, we would like to stress that the factorization property we have proven here holds
only for conformal field theories that satisfy the unitarity bound for the dimensions of operators.
Clearly, all unitary CFT's have this property, but it is possible to conceive of non-unitary CFT's
which also satisfy it. Without the unitarity bound's constraint on operator dimesions, however,
various operators we have not considered could appear in all the charge conservation identities we
have written. These operators make it possible to construct theories where the three-point
function of the stress tensor decomposes as a nontrivial superpositions of the bosonic, fermionic,
and tensor sectors. For example, we show in appendix \ref{5d-maxwell} that the free
five-dimensional Maxwell field is a non-unitary conformal field theory whose stress tensor
decomposes into a superposition of all three sectors. 

Returning to the main argument, we may now obtain all the other correlation functions, we may expand
equation (\ref{e:vvv}) to higher orders in $x_1-x_2$, and use the correlation functions obtained at
lower orders to fix the ones that appear at higher orders. For example, at second order in
$x_1-x_2$, we have:
\begin{multline}
v_{--} = (x_1-x_2)^2 \left(:\pd^2
F_{-\alpha}\Big(\frac{x_1+x_2}{2}\right)F_{-\alpha}\left(\frac{x_1+x_2}{2}\right):\\
+:\pd
F_{-\alpha}\left(\frac{x_1+x_2}{2}\right)\pd F_{-\alpha}\left(\frac{x_1+x_2}{2}\right):\Big),
\end{multline} 
and $V_{--}$ contains terms involving only the spin $2$, $3$, and $4$ currents. 
Using our answers for $\vev{222}$ and our knowledge that $\vev{223} = 0$, we can then
fix $\vev{224}$ to agree with the free field theory. This procedure recursively fixes all the 
correlators in the free tensor sector. The argument flows identically for the free bosonic and free
fermionic sectors, except that the zeroth order term will not fix $\vev{222}$, but some lower-order
current. For example, in the bosonic theory, the zeroth order term will fix $\vev{000}$, and one
will need to carry out the power series expansion to higher orders in order to fix the correlators
of the higher-spin conserved currents.

Then, one could consider correlation functions that have indices set to values other than minus.
This works in exactly the same way, since the operator product expansion of two currents with minus
indices will contain currents with other indices. This has the effect of doubling the number of
bilocals required to build a correlation function, since we need to take an extra OPE to fix the index
structure. Thus, an $n$-point function with non-minus indices can be fixed from $2n$ bilocals.
Thus, we have fixed every correlation function from currents at appear in successive OPE's of two stress
tensors, including those of every higher-spin current.

The last thing we will argue is that the normalization of the correlation functions matches the
normalization for some free theory. For example, in the theory of $N$ free bosons, the 
two-point function of $\sum_{i=1}^N :\phi_i\phi^*_i:$ will have overall coefficient $N$. 
The same is true for the fermionic and tensor cases. One might wonder if the overall coefficient
$\tilde{N}$ of the quasi-bilocal could be non-integer, which would imply that it could not coincide
with any theory of $N$ free bosons. We will now argue that this is not possible. We start with the
bosonic case, which works similarly to the argument presented in \cite{Maldacena:2011jn}:

In a theory of $N$ free bosons, consider the operator
\begin{equation}
\mathcal{O}_{q,free} = \delta^{[i_1, \dots,i_q]}_{[j_1,\dots,j_q]}
(\phi^{i_1}\pd\phi^{i_2}\dots\pd^{q-1}\phi^{i_q})
(\phi^{j_1}\pd\phi^{j_2}\dots\pd^{q-1}\phi^{j_q})
\label{e:oq-def}
\end{equation}
Here, $\delta$ is the totally antisymmetric delta function that arises from a partial contraction of
$\epsilon$ symbols:
\begin{equation}
\delta^{[i_1,\dots,i_q]}_{[j_1,\dots,j_q]} \propto 
\epsilon^{i_1\dots,i_q, i_{q+1}\dots i_N}\epsilon_{j_1\dots,j_q, i_{q+1}\dots i_N} 
\end{equation}
We claim that in the full theory, there exists an operator $\mathcal{O}_q$ in the full
theory whose correlation functions coincide with the correlation
functions of $\mathcal{O}_{q,free}$ in the free theory. The proof of this is given in Appendix 
\ref{oq-proof}.

Consider the norm of the state that $\mathcal{O}_q$ generates. This is computed by the two
point function $\vev{\mathcal{O}_q\mathcal{O}_q}$. It is obvious from the definition of
$\mathcal{O}_q$ that it arises from the contraction of $q$ bilocal fields, so this correlator is a
polynomial in $N$ of order $q$. The antisymmetry of the totally antisymmetric function in the
definition of $\mathcal{O}_{q,free}$ enforces that the correlation function vanishes at $q > N$. So
we know all the roots of the polynomial, and hence the correlation function is proportional to
$N(N-1)\dots(N-(q-1))$. Now, consider an analytic continuation of this correlator to non-integer
$\tilde{N}$. By taking $q = \lfloor N \rfloor + 2$, we find that this product is negative, which is
impossible for the norm of a state. Since the correlators of $\mathcal{O}_q$ are forced to agree
with the correlators of some operator in the full CFT, we conclude that the normalization
$\tilde{N}$ of the scalar quasi-bilocals must be an integer. 

The same argument can be ran in the tensor case for an operator defined similarly:
\begin{equation}
\mathcal{O}_q = \delta^{[i_1, \dots,i_q]}_{[j_1,\dots,j_q]}
(F_{-\{\alpha_1\}}^{i_1}\pd F_{-\{\alpha_2\}}^{i_2}\dots \pd^{q-1}F_{-\{\alpha_q\}}^{i_q})
(F_{-\{\alpha_1\}}^{j_1}\pd F_{-\{\alpha_2\}}^{j_2}\dots\pd^{q-1}F_{-\{\alpha_q\}}^{j_q})
\end{equation}
We again conclude that the normalization constant $\tilde{N}$ must be an integer.

The construction in the fermionic case is somewhat simpler. We know $j_2$ appears in $F_-$, and we
can define an operator $\mathcal{O}_q = (j_2)^q$ by extracting the term in the operator product
expansion of $q$ copies of $j_2$ whose correlation functions coincide with the free fermion operator
$(j_2)^q_{\text{free}}$. In the theory of $N$ free fermions, $j_2 = \sum_i
(\pd\psi_i)\gamma_-\bar{\psi_i} - \psi_i\gamma_-(\pd\bar\psi_i)$, where here $i$ is the flavor index
for the $N$ fermions. By antisymmetry of the fermions, we know that $\mathcal{O}_q$ will be zero if
$q \ge N$. Then, as in the bosonic case, we can consider the norm of the state that $\mathcal{O}_q$
generates, which is computed by $\vev{\mathcal{O}_q\mathcal{O}_q}$, and the rest of the argument
runs as before. Thus, the normalization $\tilde{N}$ of the fermionic bilocals must be an integer.

It is worth noting the relationship between this result and one of the primary motivations for studying
higher-spin CFT's - holographic dualities involving Vasiliev gravity in an anti-de Sitter space.
As mentioned earlier, it has been conjectured that Vasiliev gravity is conjectured to be dual to a
theory of $N$ free scalar fields in the $O(N)$ singlet sector. This implies a relationship between
the vacuum energy of Vasiliev gravity at tree-level and the free energy of a scalar field, namely, that
$F_{\text{Vasiliev}}/G_N \sim NF_{\text{scalar}}$, where $G_N$ is the Newton constant. Our result
implies that this normalization constant $N$, and therefore, the Newton constant $G_N$ is quantized
in the Vasiliev theory in any dimension.

It must be noted, however, that we cannot claim that this quantization can be seen perturbatively in
$N$. Recent work of Giombi and Klebanov \cite{Giombi:2013fka} have shown that the one-loop
correction to the vacuum energy of minimally coupled type A Vasiliev gravity in anti-de Sitter
background does not vanish as expected. This was interpreted as a shift of $N \rightarrow N-1$ in
the tree-level calculation of the vacuum free energy. Our result cannot predict such a shift or any
other $1/N$ corrections that appear in higher orders in perturbation theory. We claim only that the
exact result, after summing all loop corrections, must be quantized.

\section{Discussion and conclusions}
In this paper, we have shown that in a unitary conformal field theory in $d>3$ dimensions with
a unique stress tensor and a symmetric conserved current of spin higher than $2$, the three-point
function of the stress tensor must coincide with the three-point function of the stress tensor in
either a theory of free bosons, a theory of free fermions, or a theory of free
$\frac{d-2}{2}$-forms. This implies that all the correlation functions of symmetric currents of the
theory coincide with the those in the corresponding free field theory. 

Our technique was to use a set of appropriate lightcone limits 
to transform the data of certain key Ward identities into simple
polynomial equations. Even though we could not directly solve for the coefficients in these
identities like in three dimensions, we were nevertheless able to show that the only solution these
Ward identities admit is the one furnished by the appropriate free field theory. This was the key
step that allowed us to defined bilocal operators which were used to show that the three-point
function of the stress tensor must agree with a free field theory. This in turn fixed all the other
correlators of the theory to agree with those in the same free field theory. These results can be
understood as an extension of the techniques and conclusions of \cite{Maldacena:2011jn} from three
dimensions to all dimensions higher than three.

We stress that our classification into the bosonic, fermionic, and tensor free field theories
depends somewhat sharply on our assumption that a unique stress tensor exists. Other free field theories
with higher spin symmetry exist in $d>3$ dimensions, such as a theory of free gravitons. This
theory, however, does not have a stress tensor, and we make no statement about how the correlation
functions of such theories are constrained, and analogously for theories with many stress tensors.
On the other hand, we may consider the possibility of multiple stress tensors. It was argued in
\cite{Maldacena:2011jn} that the result holds if there are two stress tensors instead of just one.
This argument carries over to our result totally unchanged, and so our result also holds in the case
of two stress tensors. We do not comment on the possibility of more than two stress tensors.

Moreover, we have not computed correlation functions or commutators for asymmetric currents and
charges. In \cite{Boulanger:2011se}, it was shown that if one considers the possible algebras of
charges in theories that contain asymmetric currents in four dimensions, a one-parameter family of
algebras exists. This may suggest the existence of nontrivial higher-spin theories, though our
result indicates that at least the subalgebra generated by the symmetric currents must agree with
free field theory.

We also stress that the tensor structure is not well understood in all dimensions. In even
dimensions, it corresponds to the theory of a free $\frac{d-2}{2}$-form field, which does not exist
in odd dimensions. In odd dimensions, the structure may not exist, and even it does, there may not 
exist a conformal field theory which realizes it. Our argument only tells us that if there is a 
solution of the conformal and higher-spin Ward identities corresponding to this structure, then
it is unique. If the structure exists, we only know for a fact that it contains an infinite tower of
higher-spin currents for $d\ge 7$ and in this case, the theory, if it exists, has the correlation
functions we claimed.  In $d=5$, it is not known if all the higher-spin currents must be present.
Assuming they are present, our results also flow through in $d=5$.  Even then, the tensor structure
in odd dimensions could fail to have a good microscopic interpretation for many other reasons. For
example, the four-point function of the stress tensor in this sector may not be consistent with the
operator product expansion in the sense that it may not be decomposable as a sum over conformal
blocks - i.e.~it may be possible to continue all the correlation functions to odd dimensions, but
not the blocks. We have not explored this question.

\acknowledgments
We would like to thank J.~Maldacena for countless helpful discussions and invaluable
guidance on this project, A.~Zhiboedov for many useful discussions and
collaboration on parts of this paper, and J.~Penedones for his help understanding the role of the
collinear conformal group in the fermionic and tensor cases. We would also like to thank
E.~Skvortsov and D.~Ponomarev for their help in understanding their recent paper. 

The work of VA was supported by the National Science Foundation under Grant No. PHY-0756966.
The work of KD was supported by the National Science Foundation Graduate Research Fellowship under
Grant No.~DGE 1148900.

\appendix\label{appendix}
\section{Form factors as Fourier transforms of correlation functions}\label{fourier-transforms}
In this appendix, we will explicitly calculate the Fourier-transformed, lightcone-limit three-point 
functions $F^b_s$, $F^f_s$, and $F^v_s$ cited in section \ref{lightcone-limits}.
Let's start with the bosonic case. We want to compute the relevant Fourier transformation of the
three-point function $\vev{\phi(x_1)\phi^*(x_2)j_s(x_3)}$. The explicit form of $j_s(x_3)$ is given
in \cite{Anselmi:1999bb} as:
\begin{equation}
j_s = \sum_{k=0}^s c_k\pd^k\phi\pd^{s-k}\phi^*, \,\,\,\, c_k =
\frac{(-1)^k}{2}\frac{\binom{s}{k}\binom{s+d-4}{k+\frac{d}{2}-2}}{\binom{s+d-4}{\frac{d}{2}-2}}
\end{equation}
Wick's theorem and translation invariance of the correlatiors yields that::
\begin{align}
\vev{\phi(x_1)\phi^*(x_2)j_s(x_3)} 
&= \sum c_i (\pd_3^i\vev{\phi(x_1)\phi^*(x_3)}) (\pd_3^{s-i}\vev{\phi(x_3)\phi^*(x_2)})\\
&= \sum c_i (\pd_1^i\vev{\phi(x_1)\phi^*(x_3)}) (\pd_2^{s-i}\vev{\phi(x_3)\phi^*(x_2)})
\end{align}
Then, we may Fourier transform term by term with respect to $x_1^-$ and $x_2^-$. Recalling that the
propagator of a scalar field is $(x^2)^{\frac{2-d}{2}}$ and that in the lightcone limit, $x_1^+ =
x_2^+$ and $\vec{y}_1 = \vec{y}_2$, we obtain:
\begin{align}
&\pd_1^{s-i}\pd_2^i\vev{\phi(x_1)\phi^*(x_3)}\vev{\phi(x_3)\phi^*(x_2)} \nonumber \\
&\longrightarrow i^s(p_1^+)^{s-i}(p_2^+)^i \int dx_1^-dx_2^- e^{ip_1^+x_1^-}e^{ip_2^+x_2^-}
\frac{1}{\left(x_{13}^+x_{13}^- + \vec{y}_{13}^2\right)^{\frac{d-2}{2}}}
\frac{1}{\left(x_{23}^+x_{23}^- + \vec{y}_{23}^2\right)^{\frac{d-2}{2}}}\\
&= \frac{i^s(p_1^+)^{s-i}(p_2^+)^i}{(x_{13}^+)^{d-2}} \int dx_1^-dx_2^- e^{ip_1^+x_1^-}e^{ip_2^+x_2^-}
\frac{1}{\left(x_{13}^- + \frac{\vec{y}_{13}^2}{x_{13}^+}\right)^{\frac{d-2}{2}}}
\frac{1}{\left(x_{23}^- + \frac{\vec{y}_{13}^2}{x_{13}^+}\right)^{\frac{d-2}{2}}}\\
&= \frac{i^s(p_1^+)^{s-i}(p_2^+)^i}{(x_{13}^+)^{d-2}} 
\left(\int dx_1^- e^{ip_1^+x_1^-} \frac{1}{\left(x_1^- - \bar{x}\right)^{\frac{d-2}{2}}}\right)
\left(\int dx_2^- e^{ip_2^+x_2^-} \frac{1}{\left(x_2^- - \bar{x}\right)^{\frac{d-2}{2}}}\right)
\end{align}
Here, we have defined $\bar{x} = x_3^- - \frac{\vec{y}_{13}^2}{x_{13}^+}$.
Depending on the parity of $d$, each integral has either a pole of order $\frac{d-2}{2}$ 
at $\bar{x}$ or a branch point at $\bar{x}$. Our prescription for evaluating this integral is as
follows: First, we shift $x_1^-$ and $x_2^-$ by $\bar{x}$ so that the singularity is at $0$, and
then we will move move the singularity from $0$ to $\text{sign}(p)i\epsilon$. Then, the integral can be
evaluated by Schwinger parameterization. For example, suppose $p_1^+$ and $p_2^+$ are positive. 
Following our procedure, the $x_1$ integral becomes:
\begin{align}
\int_{-\infty}^\infty dx_1^- e^{ip_1^+x_1^-} \frac{1}{\left(x_1^- - \bar{x}\right)^{\frac{d-2}{2}}}
&= e^{ip_1^+\bar{x} + p_1^+\epsilon}\int_{-\infty}^\infty dy
e^{ip_1^+y}\frac{1}{(y-i\epsilon)^{\frac{d-2}{2}}}\\
&= e^{ip_1^+\bar{x} + p_1^+\epsilon}\int_{-\infty}^\infty dy 
\int_0^\infty ds \frac{i}{\Gamma(\frac{d-2}{2})}e^{ip_1^+y}s^{\frac{d-4}{2}} e^{-is(y-i\epsilon)}\\
&= \frac{ie^{ip_1^+\bar{x} + p_1^+\epsilon}}{\Gamma(\frac{d-2}{2})}
\int_0^\infty ds 2\pi\delta(s-p_1^+) e^{ip_1^+y}s^{\frac{d-4}{2}} e^{-s\epsilon}\\
&= \frac{2\pi i e^{ip_1^+\bar{x}}}{\Gamma(\frac{d-2}{2})} (p_1^+)^{\frac{d-4}{2}}
\end{align}
This function is indeed nonsingular, as required. The $x_2$ integral has exactly the same form, and
so gives the same answer. Hence, we obtain that the Fourier transform of $\vev{\phi\phi^*j_s}$ is
indeed proportional to $\sum c_i (p_1^+)^i(p_2^+)^{s-i}$, where the proportionality factor is a
nonsingular function. The, noting that the coefficients $c_i$ are the coefficients for the 
hypergeometric function with appropriate arguments, we obtain the answer cited in the text:
\begin{equation}
F^b_s \equiv \vev{\underline{\phi\phi^*}j_s} \propto
(p^+_2)^s{_2}F_1\left(2-\frac{d}{2}-s,-s,\frac{d}{2}-1,p^+_1/p^+_2\right)
\end{equation}
The fermionic and tensor cases can be tackled in exactly the same way. 
There are only two differences. First, the
propagator in the free fermion and free tensor theories are $(x^2)^{\frac{1-d}{2}}$ and
$(x^2)^{\frac{-d}{2}}$, respectively, as compared with the free scalar propagator
$(x^2)^{\frac{2-d}{2}}$. Second, the coefficients in the expression for $j_s$ are different, as can
be checked from the expressions in \cite{Dobrev:1981qc} \cite{Dobrev:1982xw} or in \cite{Anselmi:1999bb}. The end result is that the arguments of the
hypergeometric function are different in the way claimed in the text.

\section{Uniqueness of three-point functions in the tensor lightcone limit}\label{vector-uniqueness}
Our goal in this section is to show that the free tensor solution for the lightcone limit of
three-point functions explained in section \ref{lightcone-limits} is indeed unique, at least in the
lightcone limit.

Note that Lorentz symmetry constrains the propagator of spin $j$ field to be of the form
\begin{equation}
\vev{\psi_{-j}(x)\bar\psi_{-j}(0)}\propto (x^+)^{2j}.
\end{equation}
Generically, according to \cite{Zhiboedov:2012bm}, the most generic conformally invariant expression
one can write down for a three-point function of symmetric conserved currents with tensor-type 
coordinate dependence is:
\begin{multline}
\vev{j_{s_1}j_{s_2}j_{s_3}}=\frac{1}{x_{12}^{d-2} x_{23}^{d-2}x_{13}^{d-2}}
\sum\limits_{a,b,c}\Big((\Lambda_1^2\alpha_{a,b,c}+\Lambda_2\beta_{a,b,c})\left(P_{12}
P_{21}\right){}^a Q_1^b \\
\left(P_{23} P_{32}\right){}^c \left(P_{13} P_{31}\right){}^{-a-b+s_1} Q_2^{-a-c+s_2}
Q_3^{a+b-c-s_1+s_3}\Big)
\end{multline}
where the $\alpha_{a,b,c}$ and $\beta_{a,b,c}$ are free coefficients, and the $\Lambda_i$ are defined as: 
\begin{align}
\Lambda_1&= Q_1Q_2Q_3+\left[ Q_1 P_{23}P_{32}+Q_2P_{13}P_{31}+Q_3P_{12}P_{21}\right],
\\
\Lambda_2&=8P_{12}P_{21}P_{23}P_{32}P_{13}P_{31}.
\end{align}
Here, the $P$ and $Q$ invariants are defined as in \cite{Giombi:2011rz} and \cite{Todorov:2012xx}.
However, for the choice of polarization vector $\epsilon^\mu=\epsilon^-$ there is a nontrivial relation:
\begin{equation}
\Lambda_2\big|_{\epsilon^\mu_i=\epsilon^-}=-2\Lambda_1^2\big|_{\epsilon^\mu_i=\epsilon^-}, \quad \Lambda_1\big|_{\epsilon_i^\mu=\epsilon^-}=\frac{1}{4}\frac{x^+_{12}x^+_{23}x^+_{13}}{x^2_{12}x_{23}^2x_{13}^2} (\epsilon^-)^3.
\end{equation}
Therefore, in the case $\epsilon^\mu=\epsilon^-$ the expression for this three-point function
greatly simplifies. Instead of having two sets of undetermined coefficients $c_a$ and $d_a$, one
can combine the $\Lambda_i$'s into a single prefactor $\alpha_1 \Lambda_1^2+\alpha_2\Lambda_2$,
where the $\alpha_i$ are arbitrary and can be chosen to be convenient; to produce exact agreement
with the canonically normalized free-tensor theory, we will choose $\alpha_1 = 1$ and $\alpha_2 =
\frac{1}{2(d-2)}$. Now, we take the lightcone limit, which corresponds to the point
where
\begin{equation}
P_{23}P_{32}=0,\quad Q_1=-\left( \frac{P_{13}P_{31}}{Q_3}+\frac{P_{12}P_{21}}{Q_2}\right)
\end{equation}
in $P_{ij}, Q_i$ space. Then, the three-point function reduces to
\begin{equation}
\vev{j_{s_1}\underline{j_{s_2}j_{s_3}}_t}=\frac{\Lambda_1^2+\Lambda_2/(2(d-2))}{x_{12}^{d-2} x_{23}^{d-2}x_{13}^{d-2}} \sum\limits_{a=0}^{s_1-2}c_a\left( P_{12}P_{21}\right)^a \left( P_{13}P_{31}\right)^{s_1-2-a} Q_{2}^{s_2-a}Q_{3}^{S_3-s_1+a}, 
\end{equation}
Now, the $c_a$ can be fixed demanding that all currents are conserved. The result is given by the
following recurrence relation, with $c_0 = 1$:
\[ \frac{c(a+1)}{c(a)} = \frac{(s_1-2-a)(s_1 + \frac{d-4}{2}-a)(s_2+a+\frac{d-2}{2})}
{(a+1)(a+\frac{d-2}{2}+2)(s_1+s_3+\frac{d-4}{2}-2-a)} \]
This solution exactly coincides with the free tensor solution, as required.
\section{Uniqueness of $\vev{s22}$ for $s \ge 4$}\label{uniqueness-of-22s}
Define
\begin{equation}
\vev{j_{s_1}j_{s_2}j_{s_3}}=\frac{\vev{\vev{j_{s_1}j_{s_2}j_{s_3}}}}{{x_{12}}^{d-2}{x_{23}}^{d-2}{x_{13}}^{d-2}}.
\end{equation}
Using the previous defined $V$ and $H$ conformal invariants, we can write the most general expression for a conformally invariant
correlation function as follows:
\begin{multline}
\vev{\vev{j_sj_2j_2}} = V_1^{s-4} \Big[ a_1 H_{1,2}^2 H_{1,3}^2+ a_2 \left(V_1 V_2 H_{1,2} H_{1,3}^2+V_1
V_3 H_{1,2}^2 H_{1,3}\right) + a_3 V_1^2 H_{1,2} H_{1,3} H_{2,3}+
\\
+a_4 \left( V_1^2 V_3^2 H_{1,2}^2+V_1^2V_2^2 H_{1,3}^2 \right) +a_5 V_1^2V_2 V_3 H_{1,2} H_{1,3}+
\\
+a_6 \left( V_1^3V_2 H_{1,3} H_{2,3}+ V_1^3V_3 H_{1,2} H_{2,3}\right)+a_7 \left( V_1^3V_2 V_3^2
H_{1,2}+V_1^3V_2^2 V_3 H_{1,3}\right)+
\\
a_8 V_1^4 H_{2,3}^2+a_9V_1^4 V_2 V_3 H_{2,3}+a_{10} V_1^4V_2^2 V_3^2 \Big].
\end{multline}
The coefficients can be solved by imposing charge conservation. For example, in $d=4$ we obtain:
\begin{align}
a_1&= -\frac{a_7 (s-3) (s-1) (s-2)^2}{32 (s+1) (s+4)}+\frac{a_4 (s-5) (s-3) s (s-2)}{8 (s+1) (s+4)}+\frac{a_5 (s-3) (s-2)}{8 (s+4)},
\\
a_2&= -\frac{a_4 (s-2)^2}{s+4}+\frac{a_7 (s-1) (s-2)}{4 (s+4)}-\frac{a_5 (s-2)}{2 (s+4)},
\\
a_3&= -\frac{8 a_4 \left(s^2-3 s-1\right)}{(s+1) (s+4)}+\frac{a_5 (s-8)}{2 (s+4)}+\frac{a_7 (s-1) (2 s-1)}{(s+1) (s+4)},
\\
a_6&= \frac{12 a_4 (s-2)}{(s-1) (s+4)}+\frac{6 a_5}{(s-1) (s+4)}+\frac{a_7 (s-2)}{2 (s+4)},
\\
a_8&= \frac{a_7 (s-2) \left(s^2+11 s-2\right)}{4 s (s+1) (s+4)}-\frac{6 a_4 (s-5)}{(s+1) (s+4)}+\frac{a_5 (s-2)}{s (s+4)},
\\
a_9&= \frac{a_7 \left(s^2+8 s-8\right)}{s (s+4)}-\frac{24 a_4 (s-2)}{(s-1) (s+4)}+\frac{4 a_5 (s-2) (s+2)}{(s-1) s (s+4)},
\\
a_{10}&= \frac{a_7 \left(s^2+8 s+4\right)}{s (s+4)}-\frac{24 a_4 (s+1)}{(s-1) (s+4)}+\frac{4 a_5 (s+1) (s+2)}{(s-1) s (s+4)}.
\end{align}
Therefore, $\vev{\vev{j_sj_2j_2}}_t$ depends only on three parameters. The bosonic light-cone limit
of this function is zero if 
\begin{equation}
a_5= \frac{a_7 (s-2) (s-1)}{4 (s+1)}-\frac{a_4 (s-5)
s}{s+1}.
\end{equation}
The fermionic light-cone limit of this function is also zero if 
\begin{equation}a_4= \frac{a_7}{4}.\end{equation}
Therefore, $\vev{\vev{s22}}_t$ depends only on one parameter or in other words it is unique up to a
rescaling\footnote{In \cite{Stanev:2012nq} it was proven that there are only three structures for
$\vev{\vev{22s}}$ in d=4.} \begin{equation}
\vev{\vev{j_sj_2j_2}}_t \propto V_1^{s-2}\Big[H_{12}^2 V_3^2+\left(H_{23} V_1+V_2 \left(H_{13}+2 V_1 V_3\right)\right){}^2+H_{12} \left(H_{13}+2 V_1 V_3\right) \left(H_{23}+2 V_2 V_3\right)\Big], 
\end{equation}
In arbitrary dimension $d>3$, the full expression is:
\begin{align}
\vev{\vev{j_sj_2j_2}}_t &= V_1^{s-2} \left[\left(H_{23} V_1+H_{13} V_2+H_{12} V_3+2 V_2 V_3
V_1\right){}^2+\frac{2}{(d-2)} H_{12}H_{13} H_{23}\right]\nonumber\\
&=V_1^{s-2}\left[ \Lambda_1^2+\frac{1}{2(d-2)}\Lambda_2\right].
\end{align}
This formula coincides with the expression that was proposed in \cite{Zhiboedov:2012bm}, and we have
proven that this structure is unique.
\section{Transformation properties of bilocal operators under $K_-$}\label{primary-proof}
In this appendix, we will prove \ref{e:fermion-primary} and \ref{e:vector-primary} by computing
the action of a finite conformal transformation on them. The same results can be proven using 
the infinitesimal transformations, e.g.~by using equation (3) of
\cite{Weinberg:2010fx} and supplying the correct representation matrices for the Lie algebra of the
Lorentz group. One can then check that the two computations agree by expanding our results to first order in 
$b$ (remembering that only $b^-$ is nonzero for $K_-$).
\subsection{Fermionic case}
Consider a special conformal transformation
\begin{equation}
x^\mu\rightarrow y^\mu =\frac{x^\mu-b^\mu x^2}{1-2(b\cdot x)+b^2x^2}
\end{equation}
Under $K_-$, the parameter $b^\mu = b^- \delta^\mu_-$. We know that $F_-$ has the same
transformation properties as the contraction of free fields $\bar{\psi}\gamma_-\psi$ on the lightcone.
Since $K_-$ sends the lightcone into the lightcone, $V_{--}$ transforms the same way as 
$\bar{\psi}\gamma_-\psi$ under $K_-$. Using the well-known expression for the finite conformal
transformation of a Dirac spinor (e.g. \cite{Jackiw:2011vz})
\begin{align}
\psi(y)&=\left| \frac{\pd y}{\pd x}\right|^{\Delta-1/2}(1-b_\mu x_\nu\gamma^\mu\gamma^\nu)\psi(x)\\
\bar\psi(y)&=\left| \frac{\pd y}{\pd x}\right|^{\Delta-1/2}\bar\psi(x)(1-b_\mu x_\nu\gamma^\nu\gamma^\mu)
\end{align}
we may therefore compute: 
\begin{align}
F_-(\underline{y_1,y_2}) &\sim \bar\psi(y_1)\gamma^+\psi(y_2)\\ 
&= \left| \frac{\pd y_1}{\pd x_1}\right|^{\Delta-1/2}\left| \frac{\pd y_2}{\pd x_2}\right|^{\Delta-1/2}\bar\psi(x_1)(1-b_\mu (x_1)_\nu\gamma^\nu\gamma^\mu)\gamma^+(1-b_\mu (x_2)_\nu\gamma^\mu\gamma^\nu)\psi(x_2)\\
&=\left| \frac{\pd y_1}{\pd x_1}\right|^{\Delta-1/2}\left| \frac{\pd y_2}{\pd x_2}\right|^{\Delta-1/2}\bar\psi(x_1)\nonumber
\\
&\times[\gamma^+-b_+(x_1)_\nu \gamma^\nu\gamma^+\gamma^+-\gamma^+b_+(x_2)_\nu\gamma^+\gamma^\nu+b_+(x_1)_\nu \gamma^\nu\gamma^+\gamma^+b_+(x_2)_\mu\gamma^+\gamma^\mu]\psi(x_2)
\\
&=\left| \frac{\pd y_1}{\pd x_1}\right|^{\Delta-1/2}\left| \frac{\pd y_2}{\pd x_2}\right|^{\Delta-1/2} \bar\psi(x_1)\gamma^+\psi(x_2)\\
&=\Omega^{d/2-1}(x_1)\Omega^{d/2-1}(x_2)F_-(\underline{x_1,x_2})
\end{align}
The cancellations occur because $\gamma^+\gamma^+ = \eta^{++} = 0$. This is exactly equation \ref{e:fermion-primary}.
\subsection{Tensor case}
We'll start with the four-dimensional case for ease of notation and then at the end, we'll describe
how one can generalize the computation to all dimensions. Consider a special conformal transformation
\begin{equation}
x^\mu\rightarrow y^\mu =\frac{x^\mu-b^\mu x^2}{1-2(b\cdot x)+b^2x^2}
\end{equation}
Under $K_-$, the parameter $b^\mu = b^- \delta^\mu_-$. We know that $V_{--}$ has the same
transformation properties as the contraction of free fields $F_{-\mu} F_-^\mu$ on the lightcone.
Since $K_-$ sends the lightcone into the lightcone, $V_{--}$ transforms the same way as 
$F_{-\mu} F_-^\mu$ under $K_-$. We therefore compute: 
\begin{align}
V_{--}(\underline{y_1,y_2}) &= \left|\frac{\pd y_1}{\pd x_1} \right|^{-\tau_F/d} 
\left|\frac{\pd y_2}{\pd x_2} \right|^{-\tau_F/d}
\frac{\pd x_1^\mu}{\pd y_1^-} \frac{\pd x_1^\nu}{\pd y_1^\alpha} 
\frac{\pd x_2^\lambda}{\pd y_2^-} \frac{\pd x_2^\rho}{\pd y_2^\beta}\eta^{\alpha\beta}
F_{\mu\nu}(x_1)F_{\lambda\rho}(x_2)
\\
&=(1-b^-x_1^+)^{\tau_F}(1-b^-x_2^+)^{\tau_F}
(1-b^- x_1^+)^2\eta^{\alpha\beta}F_{-\alpha}(x_1)F_{-\beta}(x_2)\\
&=(1-b^-x_1^+)(1-b^-x_2^+)V_{--}(\underline{x_1,x_2})\\
&=\Omega(x_1)\Omega(x_2)V_{--}(\underline{x_1,x_2})
\end{align}
In the above manipulations, $\tau_F = \Delta - s = 0$ is the twist of $F$, and in the second to last
line, we used that $x_1^+ = x_2^+$ (because the points $x_1$ and $x_2$ are $-$ separated by
hypothesis).
This immediately implies \ref{e:vector-primary} in the four-dimensional case. In general
dimensions, the twist of $F$ will not be $0$, but rather $\Delta - s = d/2 -s$, and we will have
a corresponding number of extra factors of $\pd x/\pd y$ to contract with the additional indices of $F$. 
This will make the exponent of the $\Omega$ factors equal to $\frac{d}{2}-1$ instead of $1$. 

\section{Proof that $\mathcal{O}_q$ exists}\label{oq-proof}
In this appendix, we will prove that an operator $\mathcal{O}_q$ whose correlation functions agree with the
corresponding free field operator $\mathcal{O}_{q,free}$ defined in \ref{e:oq-def} exists in the operator 
spectrum of every conformal field theory with higher-spin symmetry. As usual, we will consider the bosonic
case, since the tensor case works almost in precisely the same way. To prove our statement, we 
will show that in the free theory, for any $q \le N$
\begin{equation}
A_{q,N}(x_1,x_2, \dots, x_{q+1}) \equiv 
\vev{\underbrace{\phi^2 \phi^2 \dots \phi^2}_{\text{q copies}}\mathcal{O}_{q,free}} \neq 0
\label{e:oqnotzero}
\end{equation}
Here, $\phi^2 = \sum_i \phi_i^2$, which is known to appear in the OPE of two stress tensors. Thus,
if we prove \ref{e:oqnotzero}, then we would know that $\mathcal{O}_{q,free}$ appears in the operator
product expansion of $2q$ copies of the free field stress tensor $j_2$. Then, just as knowing the
OPE structure of products of free field stress tensors allowed us to obtain conserved currents from
products of the quasi-bilocal fields, we can obtain $\mathcal{O}_q$ in the full theory by defining
it to be the operator appearing in the operator product expansion of $2q$ copies of $j_2$ in the full
theory whose correlation functions coincide with the correlation functions of $\mathcal{O}_{q,free}$
in the free theory. Thus, it suffices to prove \ref{e:oqnotzero}.

First, note that we can immediately reduce to the $q=N$ case. This follows
from the structure of the Wick contractions in $A_{q,N}$. To see this, note that every term in
$\mathcal{O}_{q,free}$ involves exactly $q$ of the $N$ bosons, each of which appears twice for a total
of $2q$ fields. Since $\phi^2$ is bilinear in the the fields, the product of $q$ copies of $\phi^2$ will
also contain $2q$ fields. Hence, we will need all the $\phi^2$ fields to be contracted with the 
$\mathcal{O}_{q,free}$ fields in order to obtain a nonzero answer. Thus, for each term in
$\mathcal{O}_{q,free}$, none of the $N-q$ flavors not appearing in that term will contribute, and so
we can partition the terms in $A_{q,N}$ according to which of the $q$ flavors appear. Since the
correlation function is manifestly symmetric under relabelings of the $N$ $\phi_i$ fields, this
implies that each group of terms in this partition will equally contribute to the total correlation
function an amount exactly equal to $A_{q,q}$. Hence, $A_{q,N} = \binom{N}{q}A_{q,q}$, so it
suffices to show $A_{q,q}$ is nonzero.

Then, note that since $\mathcal{O}_{q,free}$ contains exactly two copies of each of the $q$ $\phi_i$ fields, 
each of the $q$ factors of $\phi^2$ must contribute a different $\phi_i$ field for the contraction to be nonzero. 
Since $\mathcal{O}_{q,free}$ is manifestly invariant under arbitrary relabelings of the $\phi_i$ fields, we may relabel
each term so that the first copy of $\phi^2$ contributes $\phi_1^2$, the second copy of $\phi^2$ contributes $\phi_2^2$
and so on. That is, we have
\begin{equation}
A_{q,q} = q!\vev{\phi_1^2(x_1)\phi_2^2(x_2)\dots\phi_q^2(x_q)\mathcal{O}_{q,free}(x_{q+1})}
\end{equation}

The correlator on the right-hand side can be easily computed by direct evaluation of the Wick contractions. To illustrate, consider
the result given by the term in $\mathcal{O}_{q,free}$ corresponding to setting the internal indices $i_k = j_k = k$ for all 
$k \in \{1, 2, \dots, q\}$. The contribution of this term is, up to a sign, given by:
\begin{equation}
\prod_{k=1}^q \pd_{q+1}^{k-1}x_{k,q+1}^{2-d}
\end{equation}
This is a rational function whose numerator is an integer. All other terms in the correlation function will be generated by 
permuting the powers of the partial derivatives that appear. Hence, each term in the overall sum will depend differently 
only each $x_i$, and the overall sum cannot cancel because the numerators have no $x_i$ dependence. Thus, the correlation
we wanted to show is nonzero is indeed nonzero, completing the proof.

\section{The free Maxwell field in five dimensions}\label{5d-maxwell}
Consider the theory of a free Maxwell field in $d$ dimensions. The Lagrangian is
\begin{equation}
\mathcal{L}= -\frac{1}{4}(F_{\mu\nu}) ^2-\frac{1}{2 \xi} (\partial A)^2
\end{equation}
where $\xi = \frac{d}{d-4}$. As was noted in \cite{ElShowk:2011gz}, this theory is a conformal
field theory with higher spin symmetry, but it is non-unitary in dimension d > 4. We claim that
this theory is an example of a conformal, non-unitary theory where the three-point function of the
stress tensor does not coincide with one of the three free structures described in the body of the
paper. This can be checked by explicit calculation. The canonical stress energy tensor is not
trace-free, and it may be improved using the procedure of \cite{Polchinski:1987dy}. The result is
\begin{multline}
T^{--} = 4\partial_+A^\rho \partial_+ A_\rho +\partial^\rho A^-\partial_\rho A^--4\partial_+A^\rho\partial_\rho A^-+4\frac{(d-4)}{d} A^- \partial_+ (\partial A) +
\\
+\frac{1}{(d-2)} \Big[ 4a ( \partial A) \partial_+ A^- + 4a \ A^-\partial_+ (\partial A)+4a\partial_+ A^\rho \partial_\rho A^- + 4a A^\rho \partial_\rho \partial_+ A^- + 
\\
+16 b A_\rho\partial_+^2 A^\rho+16b \partial_+ A_\rho\partial_+ A^\rho-2a A^-\partial^2 A^- -2a \partial^\rho A^- \partial_\rho A^-\Big]-
\\
-2\frac{(d-4)}{(d-1)} \Big [ 
\partial_+ A_\rho \partial_+ A^\rho +A_\rho \partial_+^2 A^\rho\Big],
\end{multline}
where $a=2-d/2,b=d/4-1$. Now, the three point function $\vev{T_{--}T_{--}T_{--}}$ can be evaluated
by Wick contraction, and the result can be decomposed as follows: 
\begin{equation}
\vev{T_{--}T_{--}T_{--}}=c_s \vev{T_{--}T_{--}T_{--}}_s+c_f \vev{T_{--}T_{--}T_{--}}_f+c_t \vev{T_{--}T_{--}T_{--}}_t,
\end{equation}
where
$c_s=\frac{12125}{576}, c_f=-\frac{1000}{9}, c_t=\frac{54179}{576}$. This demonstrates that
unitarity is a necessary assumption for our result; the three-point function of the stress tensor is
not the same as the result for an appropriate free field theory. It is a superposition of the three
possible structures.
\bibliographystyle{JHEP}
\bibliography{final-writeup}
\end{document}